\documentclass{aa520}

\usepackage[dvips]{graphicx}
\usepackage{epsfig}
\usepackage{txfonts}

\newcommand{\grad}{\vec{\nabla}\,}
\newcommand{\rot}{\vec{rot}\,}
\renewcommand{\div}{\mathrm{div}\,}

\begin{document}


\title{Forced oscillations in magnetized accretion disks and QPOs.}

\author{J\'er\^ome P\'etri}

\offprints{J. P\'etri}

\institute {Astronomical Institute, University of Utrecht, P.O. Box
  80000, NL-3508 TA Utrecht, The Netherlands.
\thanks{\emph{Present address:} Max-Planck-Institut f\"ur Kernphysik,
  Saupfercheckweg 1, 69117 Heidelberg, Germany.}
}

\date{Received / Accepted}

\titlerunning{Forced oscillations in MHD accretion disks}
\authorrunning{P\'etri}

\maketitle

\begin{abstract}
  
  Quasi-periodic oscillations~(QPOs) have been observed in accretion
  disks around neutron star, black hole, and white dwarf binaries with
  frequencies ranging from a few~0.1~Hz up to 1300~Hz.  Recently, a
  correlation between their low- and high-frequency components was
  discovered and fitted with a single law, irrespective of the nature
  of the compact object. That such a relation holds over 6~orders of
  magnitude strongly supports the idea that the physical mechanism
  responsible for these oscillations should be the same in all binary
  systems.
  
  We propose a new model for these QPOs based on forced oscillations
  induced in the accretion disk due to the stellar magnetic field.
  First, it is shown that a magnetized accretion disk evolving in a
  rotating nonaxisymmetric magnetic field anchored to a neutron star
  will be subject to three kinds of resonances: a corotation
  resonance, a Lindblad resonance due to a driving force, and a
  parametric resonance due to the time varying epicyclic frequencies.
  The asymmetric part of the field is assumed to contain only one
  azimuthal mode~$m\ge1$. We focus on the~$m=1$ disturbance, which is
  well studied for an inclined dipolar rotator; but our results are
  general and easily extend to~$m>1$.  However, the radial location of
  the resonances will be affected by this number~$m$.  For instance,
  with an~$m=1$ asymmetric structure, the resonances reach regions
  very close to the innermost stable circular orbit~(ISCO) and can
  account for observations of kHz-QPOs at frequencies as high
  as~1200-1300~Hz.  If we replace the dipolar by a higher order
  multipolar component, the resonance location is shifted to larger
  radius, implying lower QPO frequencies. To compare the MHD situation
  with the hydrodynamical case, we also consider an~$m=2$ component in
  the magnetic perturbation in order to prove that, at least in the
  linear regime, the conclusions in both cases are the same. In the
  second part of the paper, we focus on the linear response of a thin
  accretion disk, developing the density perturbation as the sum of
  free wave solutions and non-wavelike disturbances. In the last part,
  we show results of 2D numerical simulations of a simplified version
  of the accretion disk consisting of a column of plasma threaded by a
  vertical magnetic field. These simulations are performed for the
  Newtonian gravitational potential, as well as for a pseudo-general
  relativistic potential, which enables us to explore the behavior of
  the resonances around both rotating neutron stars and black holes.
  We found that the density perturbations are only significant in the
  region located close to the inner edge of the disk near the ISCO
  where the magnetic perturbation is maximal.  They induce
  fluctuations in the density which persist over the whole time of the
  simulations and are closely related to the spin of the magnetic
  perturbation. It is argued that the nearly periodic motion induced
  in the disk will produce high quality factor QPOs.
  
\end{abstract}

\section{INTRODUCTION}

To date, more than 20~accreting neutron stars in Low Mass X-Ray
Binaries~(LMXBs) are known to exhibit rapid variability in their X-ray
fluxes. These high frequency quasi-periodic oscillations (QPOs) show
very strong similarities in their shape, as well as in their
amplitude, and span frequencies from~300~Hz to about 1300~Hz (see van
der Klis~\cite{vanderKlis2000} for a review).

The comparison between high frequency QPOs (HFQPOs) in black holes and
neutron stars shows a scaling proportional to~$1/M_*$, where~$M_*$ is
the mass of the compact object, suggesting that these oscillations are
due to the orbital motion in the accretion disk near the innermost
stable circular orbit~(ISCO) (McClintock \&
Remillard~\cite{MacClintock2003}). Moreover, the QPO frequencies
depend on the spectral state of the source and are correlated with the
accretion rate. QPOs are observed in the tail of the power law
spectrum indicating that they are not due to thermal emission but
probably related to Comptonization of soft photons. The saturation of
the spectral index was explained by Titarchuk \&
Fiorito~(\cite{Titarchuk2004}) in the context of coronal heating and
phase transition.

Therefore, any QPO model must take not only oscillations in the
accretion disk into account but also the way photons propagate to the
observer, including photon scattering and general relativistic effects
such as beaming, Doppler shift, and aberration.  This was investigated
by Schnittman \& Bertschinger~(\cite{Schnittman2004}), who used exact
integration of the geodesic motion of hot spots in the Kerr spacetime
and looked at the harmonic content of the signal received by an
observer in the asymptotically flat spacetime.

In black hole binaries, the HFQPOs appear in pairs, as in the neutron
star system, but their frequency remains uncorrelated to the accretion
rate. They act as a voiceprint at a ratio 3:2 that favors a resonance
mechanism. For example a parametric resonance between vertical and
horizontal epicyclic frequencies was suggested by Kluzniak et
al.~(\cite{Kluzniak2004}). Another mechanism proposed by Lee et
al~(\cite{Lee2004}) is resonance due to an external driven force.

The physics of accretion onto a magnetized neutron star has been
described by Ghosh \& Lamb (\cite{Ghosh1978}, \cite{Ghosh1979a},
\cite{Ghosh1979b}). In this picture close to the star's surface, the
magnetic field channels matter to the polar caps and exerts a torque
on it, which accelerates or decelerates the neutron star depending on
the angular momentum exchange rate. This is the standard model of
magnetized accretion onto a neutron star. However, in order to explain
pulsating X-ray sources, Anzer \& B\"orner~(\cite{Anzer1980}) propose
an alternative view in which the magnetic moment of the neutron star
and the accretion disk rotation axis are perpendicular.  They also
studied the influence of the Kelvin-Helmholtz instability on such a
configuration~(Anzer \& B\"orner~\cite{Anzer1983}).

When the magnetic moment is misaligned with respect to the spin axis
of the star, Vietri \& Stella~(\cite{Vietri1998}) show that some
inhomogeneities, treated as diamagnetic blobs, could be lifted off the
disk and suffer the Lense-Thirring precession.

Lai~(\cite{Lai1999}) has identified a magnetically driven warping and
precession instability that occurs in the inner region of an accretion
disk. This occurs by interaction between the surface current density
in the disk with the stellar magnetic field and produces motions at
frequencies much lower than the orbital one. It can, therefore,
account for the low frequency QPOs~(LFQPOs), as does the usual
Lense-Thirring precession (Stella \& Vietri~\cite{Stella1998}).
Magnetic precession, combined with the relativistic dragging effect,
was applied to weakly magnetized neutron stars to explain these LFQPOS
in LMXBs, (Shirakawa \& Lai~\cite{Shirakawa2002a}). The study of the
global magnetic warping/precession process for strongly magnetized
neutron stars can also explain the millihertz QPOs in some systems,
see for example Shirakawa \& Lai~(\cite{Shirakawa2002b}).  The
Rayleigh-Taylor instability associated with Rossby waves and the
rotational splitting gives another self-consistent description of the
QPO phenomenon (Titarchuk~\cite{Titarchuk2003}).

It was realized by Kato~(\cite{Kato2001a}, \cite{Kato2001b}) that
non-axisymmetric g-mode oscillations can be trapped in a thin
relativistic disk. These are excited by the corotation resonance. The
effect of non-linear coupling between oscillations in the disk and
warp is examined by Kato~(\cite{Kato2004}). However, his study does
not include the rotation of the compact object.

Recent observations of accretion disks orbiting white dwarfs, neutron
stars, or black holes have shown a strong correlation between the low
and high frequency QPOs, LFQPOs and HFQPOs respectively, (Psaltis et
al.~\cite{Psaltis1999}, Mauche~\cite{Mauche2002}, Belloni et
al~\cite{Belloni2002}). This relation holds over more than 6~orders of
magnitude and strongly supports the idea that the QPO phenomenon is a
universal physical process independent of the nature of the compact
object. Indeed, the presence or absence of a solid surface, a magnetic
field, or an event horizon plays no relevant role in the production of
X-ray variability~(Wijnands~\cite{Wijnands2001}).  This correlation
has instead been explained in terms of the centrifugal barrier model
of Titarchuk et al.~(\cite{Titarchuk2002}) where the LFQPOs are
associated with magneto-acoustic waves, and the HFQPOs correspond to
the Keplerian motion in the disk.

In this paper we propose a new resonance in the star-disk system
arising from the response of the accretion disk to a non-axisymmetric
rotating magnetic field. The paper is organized as follows. In
Sec.~\ref{sec:EquInit}, we describe the initial equilibrium state
where the accretion disk is embedded in an axisymmetric magnetic field
anchored to the neutron star.  Furthermore, we assume that the
magnetic moment is aligned with the rotation axis of the star. We then
estimate the field perturbation induced by a small inclination of the
magnetic moment.  In Sec.~\ref{sec:AnalLin}, we look for the resonance
conditions and elucidate the meaning of the resonances. For a thin and
weakly magnetized accretion disk, assuming no warp and only motion in
the orbital plane of the disk, we investigate the precise linear
behavior of the disk which is presented in Sec.~\ref{sec:ThinDisk}. In
Sec.~\ref{sec:Simulation}, we perform 2D numerical MHD simulations of
the disk and show that some resonances persist on a long timescale
with a narrow radial extension.  This is done for Newtonian, as well
as for pseudo-general relativistic gravitational potential describing
the Kerr geometry in a Newtonian way adapted to a rotating neutron
star.  Conclusions are drawn in Sec.~\ref{sec:Conclusion}.


\section{THE INITIAL CONDITIONS}
\label{sec:EquInit}

\subsection{The stationary state}

In the equilibrium state, the accretion disk evolves in a perfectly
spherically symmetric gravitational field and an axisymmetric dipolar
magnetic field, both due to the compact object. We assume that the
magnetic moment of the star and its rotational axis are aligned. The
disk is assumed to lie in the equatorial plane of the star so that it
experiences an axisymmetric stationary field. By adopting a
cylindrical coordinate system labeled by~$(r,\varphi,z)$, we can
therefore describe all physical quantities with only a~$(r,z)$
dependence.  Neglecting the matter inflow because in the thin disk
approximation~$v_r\ll v_\varphi$ and because we only focus on
oscillations around a stationary state, the density~$\rho$,
pressure~$p$, current density~$\vec{j}$, velocity~$\vec{v}$, and
magnetic field~$\vec{B}$ at equilibrium are thus given by~:
\begin{eqnarray}
  \label{eq:GrandeurEq}
  \rho & = & \rho(r,z) \\
  p & = & p(r,z) \\
  \vec{v} & = & r \, \Omega(r,z) \, \vec{e}_\varphi \\
  \vec{B} & = & B_r(r,z) \, \vec{e}_r + B_z(r,z) \, \vec{e}_z \\
  \vec{j} & = & j_\varphi(r,z) \, \vec{e}_\varphi
\end{eqnarray}
The magnetic field has two different origins~:
\begin{itemize}
\item one component~$\vec{B}_*$ is attached to the neutron star and
  corresponds to the rotating stellar dipolar magnetic field~;
\item the second component~$\vec{B}_d$ of the magnetic field is
  induced by the fluid motion in the accretion disk and follows from
  Ampere's law~$\rot\vec{B}_d=\mu_0\,\vec{j}$.
\end{itemize}
The total magnetic field is then simply the sum of these two
components~$\vec{B}=\vec{B}_*+\vec{B}_d$. 

In the radial direction, the gravitational attraction from the compact
object~$\vec{g}$ is balanced by the centrifugal force, the Lorentz
force, and the pressure gradient, whereas in the vertical direction we
simply have hydromagnetic equilibrium~:
\begin{eqnarray}
  \label{eq:Equilibre3Dr}
   -\rho \, \frac{v_\varphi^2}{r} = \rho \, g_r - \frac{\partial p}{\partial r}
   + j_\varphi \, B_z & = & \rho \, g_r - \frac{\partial}{\partial r} \, 
   \left( p + \frac{B_z^2}{2\,\mu_0} \right) \\
  \label{eq:Equilibre3Dz}
  \rho \, g_z - \frac{\partial p}{\partial z} - j_\varphi \, B_r & = & 0
\end{eqnarray}
To specify a given equilibrium state, we prescribe the initial density
and current, or, equivalently, the magnetic field in the disk and the
strength of the dipolar magnetic field of the neutron star. Assuming a
thin accretion disk and a low~$\beta$-plasma parameter, defined as
\begin{equation}
  \label{eq:BetaPlasma}
  \beta = \frac{p}{B^2/2\,\mu_0},
\end{equation}
the pressure gradient and Lorentz forces will be negligible compared
to the gravitational force. The motion will then only deviate slightly
from Keplerian rotation. We then perturb the system by adding a small
inclination angle~$\chi$ between the magnetic and rotational axis of
the star. As a consequence, and due to the magnetic perturbation, the
disk will leave its equilibrium state and start to oscillate around
the equatorial plane. We next give the expression for the stellar
magnetic field perturbation.

\subsection{Perturbed dipolar magnetic field of a rotating star}
 
In the case of an aligned rotator,~$\chi=0$, the magnetic field is
axisymmetric. Therefore, in the disk plane, there is only a vertical
component~$B_z$. When the magnetic moment~$\vec{\mu}$ is inclined by
an angle~$\chi\ne0$ relative to rotation axis~$\vec{\Omega}_*$, an
additional magnetic component dragged by the star's rotation arises.
This perturbation is responsible for the oscillation in the disk.  We
derive its expression below. The magnetic moment can be developed into
a parallel (along~$\vec{e}_z$) and a perpendicular
(along~$\vec{e}_r^*$) component such that
\begin{equation}
  \vec{\mu} = \mu \, \sin\chi \, \vec{e}_r^* + \mu \, \cos\chi \, \vec{e}_z.
\end{equation}
where~$\vec{e}_r^*$ is a radial unitary vector corotating with the
star.  In the Cartesian coordinates of a distant observer at
rest~$(\vec{e}_x, \vec{e}_y, \vec{e}_z)$, it can be expressed as
\begin{equation}
  \vec{e}_r^* = \cos(\Omega_*\,t) \, \vec{e}_x + \sin(\Omega_*\,t) \, \vec{e}_y
\end{equation}
assuming that at~$t=0$ $\vec{e}_r^* =\vec{e}_x$.  We adopt a
particular geometric configuration corresponding to the situation
where the disk is contained in the star's equatorial plane.  We could
generalize to the case where magnetic moment, disk, and star rotation
axis are all misaligned.  However, the rotational frequency of the
magnetic perturbation, which is its most important property, would not
be affected.

The angular rotation is assumed to be equal to~$\vec{\Omega}_* =
\Omega_*\,\vec{e}_z$. The stellar dipolar magnetic field has the usual
expression derived from the vector potential
\begin{equation}
  \label{eq:Betoile}
  \vec{A}_* = \frac{\mu_0}{4\,\pi} \, \frac{\vec{\mu}\wedge\vec{r}}{r^3}.
\end{equation}
Therefore, the perturbation is~$\delta\vec{A}_* = \frac{\mu_0}{4\,\pi}
\, \mu \, \sin\chi \, \frac{\vec{e}_r^*\wedge\vec{r}}{r^3}$. It
represents a perpendicular rotator with magnetic
moment~$\mu_\perp=\mu\,\sin\chi$. Expressed in the cylindrical
coordinate system, the components of the perturbation of the magnetic
field are given by~:
\begin{eqnarray}
  \label{eq:deltaBetoileR}
  \delta B_*^r & = & \frac{\mu_0}{4 \, \pi} \, \frac{\mu \, \sin \chi}{(r^2+z^2)^{3/2}} \,
  \left[ \frac{3\,r^2}{r^2+z^2} - 1 \right] \, \cos (\varphi-\Omega_*\,t) \\
  \delta B_*^\varphi & = & \frac{\mu_0}{4 \, \pi} \, \frac{\mu \, \sin \chi}{(r^2+z^2)^{3/2}} \, 
  \sin (\varphi-\Omega_*\,t) \\
  \label{eq:deltaBetoileZ}
  \delta B_*^z & = & \frac{\mu_0}{4 \, \pi} \, \frac{\mu \, \sin \chi}{(r^2+z^2)^{3/2}} \,
  \frac{3\,r\,z}{r^2+z^2} \, \cos (\varphi-\Omega_*\,t)
\end{eqnarray}
We see that in the equatorial plane the perturbation is~$\delta B_*^z
= 0$.  Each component has a time function that depends
on~$\{\cos/\sin\}(\varphi-\Omega_*\,t)$. The nature of the driving is
therefore an~$m=1$ azimuthal mode with an excitation frequency at a
rate~$\Omega_*$. This has to be contrasted with the~$m=2$ mode and,
consequently, excitation frequency of~$2\Omega_*$ produced by the
quadrupolar gravitational field (the lowest order perturbation in
gravity for a hydrodynamical accretion disk for example).  This
situation will be studied in detail in a forthcoming paper. The nature
of the perturbation has some consequences for the derivation of the
resonance conditions, as shown in the following~Sec.\ref{sec:AnalLin}.


\section{LINEAR ANALYSIS}
\label{sec:AnalLin}

The distorted magnetic field described above will kick the disk out of
its equilibrium state. We study its linear response by treating the
asymmetric part of the magnetic field as a small perturbation of the
equilibrium state prescribed in Sec.~\ref{sec:EquInit}. We start with
the ideal MHD equations of the accretion disk with adiabatic motion
given by~:
\begin{eqnarray}
  \label{eq:DiscMHDrho}    
  \frac{\partial\rho}{\partial t} + \div (\rho\,\vec{v}) & = & 0 \\
  \label{eq:DiscMHDv}    
  \frac{\partial\vec{v}}{\partial t} + (\vec{v}\cdot\grad) 
  \, \vec{v} & = & \vec{g} - \frac{\grad p}{\rho} + \vec{j} \wedge \vec{B} \\
  \frac{D}{Dt} \left( \frac{p}{\rho^\gamma} \right) & = & 0 \\
  \label{eq:DiscMHDBind}    
  \frac{\partial\vec{B}}{\partial t} & = & \rot ( \vec{v} \wedge \vec{B} ) \\
  \label{eq:DiscMHDB}    
  \rot \vec{B} & = & \mu_0 \, \vec{j}.
\end{eqnarray}
All quantities have their usual meaning: $\rho$ density of mass in the
disk,~$\vec{v}$ the velocity of the disk, $p$ the gaseous pressure,
$\gamma$ the adiabatic index,~$\vec{g}$ the spherically symmetric
gravitational field, and~$\vec{B}$ the total magnetic field.

\subsection{Lagrangian displacement}

We then perturb the equilibrium state from Sec.~\ref{sec:EquInit} by
introducing the Lagrangian displacement~$\vec{\xi}$. Expanding to
first order we get for the Eulerian perturbations of the
magnetohydrodynamical quantities attached to the disk~:
\begin{eqnarray}
  \delta\rho & = & - \, \div (\rho \, \vec{\xi} \, ) \\
  \delta\vec{v} & = & \frac{\partial\vec{\xi}}{\partial t} + 
  \vec{v}\cdot\grad \vec{\xi} - \vec{\xi} \cdot \grad \vec{v} \\
  \delta p & = & - \, \gamma \, p \, \div \vec{\xi} - \vec{\xi}\cdot\grad p \\
  \delta\vec{B} & = & \vec{B} \cdot \grad \vec{\xi} - \vec{B} \, \div\vec{\xi} - 
  \vec{\xi} \cdot \grad \vec{B}
\end{eqnarray}
For a detailed derivation of these perturbed quantities and their
associated boundary conditions as well as for the equations described
in this section, we refer the reader to Appendix~\ref{app:Eigenvalue}.
By also making allowance for a perturbation in the neutron star
magnetic field and following the Frieman-Rotenberg analysis~(Frieman
\& Rotenberg~\cite{Frieman1960}), the Lagrangian displacement
satisfies a second order linear partial differential equation given
by~:
\begin{eqnarray}
  \label{eq:PDEXiGen}
  \rho \, \frac{D^2\vec{\xi}}{Dt^2}
  - \div\left[ \rho \, \vec{\xi} ( \vec{v} \cdot \nabla ) \vec{v} \right] - \nabla\Pi
  - \frac{1}{\mu_0} \, \vec{B} \cdot \nabla \vec{Q} & & \nonumber \\ 
  - \frac{1}{\mu_0} \, \vec{Q} \cdot \nabla \vec{B} +
  \div ( \rho \, \vec{\xi} ) ( \vec{g} + \delta\vec{g} ) - \rho \, \delta\vec{g} =
  & & \nonumber \\ 
  \frac{1}{\mu_0} \, \left[ \rot ( \vec{B} + \vec{Q} + \delta \vec{B}_* ) \wedge \delta \vec{B}_* 
    + \rot \delta \vec{B}_* \wedge ( \vec{B} + \vec{Q} ) \right]
  & &  
\end{eqnarray}
We have introduced the convective derivative by~$D/Dt = \partial_t +
\Omega\,\partial_\varphi$. The perturbation in the magnetic field is
represented by the vector~$\vec{Q} = \rot(\vec{\xi} \wedge \vec{B})$.
The scalar~$\Pi=\gamma\,p\,\div\vec{\xi} + \vec{\xi} \cdot
\vec{\nabla} p - \frac{1}{\mu_0} \, \vec{B} \cdot \vec{Q}$ corresponds
to the opposite of the total (gaseous+magnetic) pressure perturbation.
Compared to the equation obtained by Frieman \&
Rotenberg~(\cite{Frieman1960}), we have an additional term in the
Lorentz force induced by the stellar magnetic perturbation depicted by
the expression enclosed by the brackets after the $\frac{1}{\mu_0}$
factor, terms containing~$\delta \vec{B}_*$. We also allow for a
degree of freedom in the choice of an arbitrary gravitational
perturbation induced by the disk~$\delta\vec{g}$. However, for the
rest of this paper, we only focus on the consequences of an asymmetry
in the magnetic field, leaving the gravity perturbation for future
study.

We make a clear distinction between the stellar magnetic
perturbation~$\delta\vec{B}_*$ which is known at each time and given
by the rotating magnetic dipole
Eq.~(\ref{eq:deltaBetoileR})-(\ref{eq:deltaBetoileZ}) or any kind of
multipolar fluctuation associated with the star and those caused by
the current in the disk~$\delta\vec{B}$ in response to the former
perturbation.

Expressing Eq.~(\ref{eq:PDEXiGen}) in cylindrical coordinates for the
radial component leads to~:
\begin{eqnarray}
  \label{eq:XiR}
  \rho \, \frac{D^2\xi_r}{Dt^2} - 2 \, \rho \, \Omega \, \frac{D\xi_\varphi}{Dt} 
  - \frac{\partial\Pi}{\partial r} + & & \nonumber \\
  \frac{1}{\rho} \, \frac{dp}{dr} \, \div (\rho\,\vec{\xi}) + 
  \rho \, r \, \vec{\xi} \cdot \grad (\Omega^2) - & & \nonumber \\
  \frac{1}{\mu_0} \, \left[ B_r \, \frac{\partial Q_r}{\partial r} +
    \frac{B_\varphi}{r} \, \frac{\partial Q_r}{\partial \varphi} - 
    2 \, \frac{B_\varphi\,Q_\varphi}{r} + \right. & & \nonumber \\
  \left. B_z \, \frac{\partial Q_r}{\partial z} + Q_r \, \frac{\partial B_r}{\partial r} +
    \frac{Q_\varphi}{r} \, \frac{\partial B_r}{\partial \varphi} +
    Q_z \, \frac{\partial B_r}{\partial z} \right] & = & \nonumber \\
  \frac{1}{\mu_0} \,
  \left[ \delta B_*^z \, \left\{ \frac{\partial}{\partial z} 
      \left( B_r + Q_r + \delta B_*^r \right) - \frac{\partial}{\partial r} 
      \left( B_z + Q_z + \delta B_*^z \right) \right\} \right. & - & \nonumber \\
  \left. \frac{\delta B_*^\varphi}{r} \, \left\{ \frac{\partial}{\partial r} 
      \left( r \, (  B_\varphi + Q_\varphi + \delta B_*^\varphi ) 
      \right) - \frac{\partial}{\partial\varphi} 
      ( B_r + Q_r + \delta B_*^r ) \right\} \right. & + & \nonumber \\
  \left. (B_z + Q_z) \, \left( \frac{\partial\delta B_*^r}{\partial z} - 
      \frac{\partial\delta B_*^z}{\partial r} \right) - \right. & & \nonumber \\
    \left. 
    \frac{B_\varphi + Q_\varphi}{r} \, \left( \frac{\partial}{\partial r}(r\,\delta B_*^\varphi) -
      \frac{\partial\delta B_*^r}{\partial\varphi} \right) \right] & &
\end{eqnarray}
while for the azimuthal component we get~:
\begin{eqnarray}
  \label{eq:XiPhi}
  \rho \, \frac{D^2\xi_\varphi}{Dt^2} +
  2 \, \rho \, \Omega \, \frac{D\xi_r}{Dt} -
  \frac{1}{r} \, \frac{\partial\Pi}{\partial\varphi} & - & \nonumber \\
  \frac{1}{\mu_0} \, \left[ B_r \, \frac{\partial Q_\varphi}{\partial r} +
    \frac{B_\varphi}{r} \, \frac{\partial Q_\varphi}{\partial \varphi} + 
    \frac{B_\varphi\,Q_r}{r} + \right. & &  \nonumber \\
  \left.
    B_z \, \frac{\partial Q_\varphi}{\partial z} + Q_r \, \frac{\partial B_\varphi}{\partial r} +
    \frac{Q_\varphi}{r} \, \frac{\partial B_\varphi}{\partial\varphi} + 
    \frac{Q_\varphi\,B_r}{r} +
    Q_z \, \frac{\partial B_\varphi}{\partial z} \right] & = & \nonumber \\
  \frac{1}{\mu_0} \, 
  \left[ \frac{\delta B_*^r}{r} \, \left\{ \frac{\partial}{\partial r} 
      \left( r \, (  B_\varphi + Q_\varphi + \delta B_*^\varphi ) \right) - \right. \right. 
& & \nonumber \\
\left.\left.
      \frac{\partial}{\partial\varphi} ( B_r + Q_r + \delta B_*^r ) \right\} -\right. \nonumber \\
  \left.
    \delta B_*^z \left\{ \frac{1}{r} \frac{\partial}{\partial\varphi} 
      (  B_z + Q_z + \delta B_*^z ) - \frac{\partial}{\partial z} 
      (  B_\varphi + Q_\varphi + \delta B_*^\varphi )\right\} \right. & & \nonumber \\
  \left.\frac{B_r + Q_r}{r} \, \left( \frac{\partial}{\partial r}(r\,\delta B_*^\varphi) -
      \frac{\partial\delta B_*^r}{\partial\varphi} \right) - \right. & & \nonumber \\
    \left. (B_z + Q_z) \, 
    \left( \frac{1}{r} \, \frac{\partial\delta B_*^z}{\partial\varphi} -
      \frac{\partial\delta B_*^\varphi}{\partial z} \right) \right] & & 
\end{eqnarray}
and finally for the vertical oscillations we have~:
\begin{eqnarray}
  \label{eq:XiZ}
  \rho \, \frac{D^2\xi_z}{Dt^2} - \frac{\partial\Pi}{\partial z} -
  \frac{1}{\mu_0} \, \left[ B_r \, \frac{\partial Q_z}{\partial r} + 
    \frac{B_\varphi}{r} \, \frac{\partial Q_z}{\partial \varphi} +
    B_z \, \frac{\partial Q_z}{\partial z} + \right. & & \nonumber \\
  \left. Q_r \, \frac{\partial B_z}{\partial r} + 
    \frac{Q_\varphi}{r} \, \frac{\partial B_z}{\partial \varphi} +
    Q_z \, \frac{\partial B_z}{\partial z} \right] =
  - \div (\rho\,\vec{\xi}) ) \, g_z & + & \nonumber \\
  \frac{1}{\mu_0} \, \left[ \delta B_*^\varphi \left\{ \frac{1}{r} \frac{\partial}{\partial\varphi} 
      (  B_z + Q_z + \delta B_*^z ) - \right. \right. & & \nonumber \\
\left. \left. \frac{\partial}{\partial z} 
      (  B_\varphi + Q_\varphi + \delta B_*^\varphi )\right\} - \right. \nonumber \\
    \delta B_*^r \, \left\{ \frac{\partial}{\partial z} 
      \left( B_r + Q_r + \delta B_*^r \right) - \frac{\partial}{\partial r} 
      \left( B_z + Q_z + \delta B_*^z \right) \right\} & + & \nonumber \\
  \left.  ( B_\varphi + Q_\varphi ) \left( \frac{1}{r} \, 
      \frac{\partial\delta B_*^z}{\partial\varphi} -
      \frac{\partial\delta B_*^\varphi}{\partial z} \right) - \right. & & \nonumber \\
    \left. 
    (B_r + Q_r) \left( \frac{\partial\delta B_*^r}{\partial z} -
      \frac{\partial\delta B_*^z}{\partial r} \right) \right] & &
\end{eqnarray}
We emphasize the fact that these expressions contain terms of the
form~$\xi_i \, \delta B_*^j$, $(i,j=r,\varphi,z)$ that are second
order with respect to the perturbation and that, therefore, could be
neglected as long as $\delta\vec{B}_*$ remains weak. But in doing so,
we suppress the parametric resonance to be studied in more detail
below. This resonance becomes relevant when the perturbed stellar
magnetic field is on the same order of magnitude as the aligned
component. Depending on the magnitude of the perturbation, which we
suppose small compared to the background field, this resonance will
develop on a timescale closely related to the amplitude of the
perturbation so should not be ignored. The terms~$\delta B_*^i\,\delta
B_*^j$ are also second order when the magnetic asymmetry is small.

Finding an analytical stability criterion for this system is a
complicated or perhaps even an impossible task. Furthermore, we cannot
apply the classical expansion in plane wave solutions leading to an
eigenvalue problem. The presence of some coefficients varying
periodically in time, like~$\delta \vec{B}_*$, prevents such a
treatment. However, the problem can be cast in a more convenient form
if we decouple the oscillations in the orbital plane from those
perpendicular to it, as is done in the following subsection.

\subsection{Detailed linear analysis of the thin disk}
\label{sec:ThinDisk}

If we assume that there is no warping in the disk and neglect the
gradient and displacement in the vertical direction, we can perform a
detailed and accurate analysis of the motions in the disk.  Indeed, by
setting~$\xi_z=0$ and~$\partial/\partial z =0$ in Eq.~(\ref{eq:XiR})
and~(\ref{eq:XiPhi}), the linearized MHD equations become much more
tractable. Moreover, we first assume a simple geometric structure in
which the disk is replaced by a column of plasma threaded by a
vertical magnetic field and then allow only magnetic perturbation in
this direction.  This situation seems far from the real system of an
oblique rotator; nevertheless, it gives a good first insight into the
oscillations and resonances arising there.

As a result, setting~$B_r=B_\varphi=\delta B_*^r = \delta B_*^\varphi
= 0$ we find for the magnetic perturbation
\begin{eqnarray}
  Q_r & = & 0 \\
  Q_\varphi & = & 0 \\
  Q_z & = & - \frac{1}{r} \, \left[ \frac{\partial}{\partial r}
  \left( r \, B_z \, \xi_r \right) + \frac{\partial}{\partial \varphi}
  \left( B_z \, \xi_\varphi \right) \right]
\end{eqnarray}
and for the opposite of the total pressure perturbation~:
\begin{equation}
  \Pi = \frac{\gamma \, p}{r} \, \left[ \frac{\partial}{\partial r}
    \left( r \, \xi_r \right) + \frac{\partial\xi_\varphi}{\partial \varphi} 
  \right] + \xi_r \, \frac{\partial p}{\partial r} - \frac{B_z\,Q_z}{\mu_0}.
\end{equation}
We seek solutions by writing each {\it perturbation}, such as the
components of the {\it perturbed} magnetic field~$\delta\vec{B}$ (i.e.
those of $\vec{Q}$), those of the Lagrangian displacement~$\vec{\xi}$,
those of the {\it perturbed} velocity~$\delta\vec{v}$, the {\it
  perturbed} density~$\delta \rho$, and the {\it perturbed}
pressure~$\delta p$ as
\begin{equation}
  \label{eq:DvlptX}
  X(r,\varphi,t) = {\rm Re} [\tilde{X}(r) \, e^{i(m\,\varphi-\sigma\,t)}],
\end{equation}
where~$m$ is the azimuthal wavenumber and $\sigma$ the eigenfrequency of the
perturbation related to the speed pattern~$\Omega_p$ by $\sigma =
m\,\Omega_p$.  Therefore
\begin{eqnarray}
  \label{eq:Qz}
  \tilde{Q}_z & = & - \frac{1}{r} \, \frac{\partial}{\partial r}
  \left( r \, B_z \, \tilde{\xi}_r \right) - \frac{i\,m}{r} \, B_z \, \tilde{\xi}_\varphi \\
  \label{eq:Pi}
  \tilde{\Pi} & = & \rho \, c_{maz}^2 \left[ 
    \frac{1}{r} \, \frac{\partial}{\partial r} \left( r \, \tilde{\xi}_r \right) +
    \frac{i\,m}{r} \, \tilde{\xi}_\varphi \right] + \tilde{\xi}_r \, 
  \frac{\partial}{\partial r} \left( p + \frac{B_z^2}{2\,\mu_0} \right) \nonumber \\
  & & 
\end{eqnarray}
$c_{maz}^2= c_s^2 + c_{az}^2$ is the vertical fast magneto-acoustic
wave speed. The sound speed and the Alfven velocity associated with
the vertical magnetic field are respectively~:
\begin{eqnarray}
  c_s^2 & = & \frac{\gamma\,p}{\rho} \\
  c_{az}^2 & = & \frac{B_z^2}{\mu_0\,\rho}.
\end{eqnarray}
Neglecting quadratic expressions in the excitation term,
like~$Q_z\delta B_*^z$, the azimuthal Lagrangian displacement
Eq.~(\ref{eq:XiPhi}) is solved by~:
\begin{eqnarray}
  \label{eq:XiPhiSol}
  \tilde{\xi}_\varphi = - \frac{i}{\omega_*^2} \, \left[ - \frac{m}{r\,\rho} \,
    \frac{B_z\,\delta B_*^z}{\mu_0} + \left( 2 \, \Omega \, \omega +
      \frac{m}{r\,\rho} \, \frac{\partial}{\partial r} 
      \left( p + \frac{B_z^2}{2\,\mu_0} \right) \right) \, \tilde{\xi}_r + \right.
  & & \nonumber \\
  \left. \frac{m \, c_{maz}^2}{r^2} \, \frac{\partial}{\partial r} 
    \left( r \, \tilde{\xi}_r \right) \right]. & &
\end{eqnarray}
Due to the differential rotation of the disk, the
eigenfrequency~$\sigma$ appears Doppler-shifted. It is therefore
convenient to introduce the frequency~$\omega$ and a second
one~$\omega_*$ taking the speed of the magneto-sonic wave into account
by
\begin{eqnarray}
  \omega     & = & \sigma - m \, \Omega \\
  \omega_*^2 & = & \omega^2 - \frac{m^2 \, c_{maz}^2}{r^2}.
\end{eqnarray}
Strictly speaking, corotation is achieved when~$\omega_*=0$, such that
there exist two solutions. When far from the corotation points,
because of the weakly magnetized thin accretion disk
approximation~$O(c_s^2) = O(c_{az}^2) \ll O(r^2\,\omega^2)$ and
therefore~$\omega_*^2 \approx \omega^2$. Substituting the expressions
for~$\tilde{Q}_z$, Eq.~(\ref{eq:Qz}), for~$\tilde{\Pi}$,
Eq.~(\ref{eq:Pi}) and for~$\tilde{\xi}_\varphi$,
Eq.~(\ref{eq:XiPhiSol}), in the radial equation Eq.~(\ref{eq:XiR}), we
obtain to lowest order~:
\begin{eqnarray}
  \label{eq:XiRSchroed}
  c_{maz}^2 \, \left[ \frac{\partial^2\tilde{\xi}_r}{\partial r^2}
    + \frac{\partial\ln(r\,\rho\, c_{maz}^2 )}{\partial r}
    \, \frac{\partial\tilde{\xi}_r}{\partial r} \right]
  + \left( \omega^2 - \kappa_r^2
  \right) \, \tilde{\xi}_r & = & \frac{B_z}{\mu_0\,\rho} \, \frac{\partial\delta B_*^z}{\partial r}. \nonumber \\
& & 
\end{eqnarray}
The radial epicyclic frequency of a single particle as measured by an
observer in the rest frame is
\begin{equation}
  \kappa_r^2 = 4\,\Omega^2 + r\,\frac{\partial\Omega^2}{\partial r}
\end{equation}
(see Appendix~\ref{app:Eigenvalue}).  The solutions of this equation
consist of free waves corresponding to density perturbations
propagating in the disk without magnetic asymmetry related to the
homogeneous part and of non-wavelike disturbances due to this
asymmetric component related to the inhomogeneous part.

\subsubsection{Free wave solutions}

When looking for free wave solutions to~Eq.(\ref{eq:XiRSchroed}), we
can set~$\delta B_*^z=0$. Thus
\begin{equation}
  \label{eq:XiROndeLibre}
  c_{maz}^2 \, \left[ \frac{\partial^2\tilde{\xi}_r}{\partial r^2}
    + \frac{\partial\ln(r\,\rho\, c_{maz}^2 )}{\partial r}
    \, \frac{\partial\tilde{\xi}_r}{\partial r} \right] + 
    ( \omega^2 - \kappa_r^2) \, \tilde{\xi}_r = 0 .
\end{equation}
This is the generalization to MHD of the sound wave that propagates in
the hydrodynamical disk.  We just need to replace the sound
speed~$c_s^2$ by the fast magneto-acoustic speed~$c_{maz}^2$.
Introducing a new unknown~$\psi = \tilde{\xi}_r \,
\sqrt{r\,\rho\,c_{maz}^2}$, it satisfies a kind of Schr\"odinger
equation
\begin{equation}
  \label{eq:AppPsiSimp}
  \psi''(r) + V(r) \, \psi(r) = 0
\end{equation}
where the potential is given by~:
\begin{equation}
  \label{eq:AppPotentiel}
  V(r) = \frac{\omega^2 - \kappa_r^2}{c_{maz}^2}.
\end{equation}
A first guess for free wave solutions is given by the WKB expansion as
follows~:
\begin{equation}
  \label{eq:PsiSimp}
    \Psi(r) = \Phi(r) \, e^{i\int^r k(s) \, ds}.
\end{equation}
Putting this approximation into Eq.~(\ref{eq:AppPsiSimp}), the
dispersion relation for the fast magneto-acoustic waves propagating in
the orbital plane of the accretion disk is given by~:
\begin{equation}
  \label{eq:Dispersion}
  \omega^2 = \kappa_r^2 + c_{maz}^2 \, k^2.
\end{equation}
Free waves can only propagate in regions where~$\omega^2 -
\kappa_r^2=V(r)\, c_{maz}^2 \ge0$. The boundary between the
propagating and damping zones is defined by the inner and outer
Lindblad radius~$r_L^{in/out}$ defined by~$V(r_L^{in/out})=0$.  Using
the results of Appendix~\ref{app:AppSchroeApprox}, we can find a
better approximation to the solution of Eq.~(\ref{eq:AppPsiSimp})
which is valid even for~$r\approx r_L^{in/out}$. For the inner
Lindblad resonance of interest here, we introduce the
following function~$\omega_1$, writing~$r_L=r_L^{in}$~:
\begin{eqnarray}
  \omega_1(r) & = & - \left[ - \frac{3}{2} \, \int_{r_L}^r \sqrt{V(s)} \, ds \right]^{2/3}
  \mathrm{for\;} r \le r_L \\
  \omega_1(r) & = & \left[ \frac{3}{2} \, \int_{r_L}^r \sqrt{-V(s)} \, ds \right]^{2/3}
  \mathrm{for\;} r \ge r_L.
\end{eqnarray}
The function~$\psi$ is then a linear combination of the 2~linearly
independent solutions:
\begin{eqnarray}
  \label{eq:SolXi2}
  \psi_1(r) & = & \frac{Ai(\omega_1(r))}{\sqrt{|\omega_1'(r)|}} \\
  \psi_2(r) & = & \frac{Bi(\omega_1(r))}{\sqrt{|\omega_1'(r)|}}.
\end{eqnarray}
Furthermore, we require the solution to remain bounded, which leads
to~$C_2=0$. Thus the solution for the Lagrangian displacement is
$\xi_r=C_1\,\psi_1(r)/\sqrt{r\,\rho\,c_{maz}^2 }$.  At the inner
boundary of the accretion disk, in accordance with the simulations
performed later on, the density perturbation should vanish. This is
expressed as~$\delta\rho=0$. To the lowest order consistent with our
approximation, the Lagrangian radial displacement~$\xi_r$ must
satisfy
\begin{equation}
  \label{eq:Boundary}
  \tilde{\xi}_r'(R_1) + \left( \left.\frac{\partial\ln(r\,\rho)}
      {\partial r}\right|_{R_1} + 
    2\,m\,\frac{\Omega}{\omega\,R_1} \right) \, \tilde{\xi}_r(R_1) = 0.
\end{equation}
This last condition determines the eigenfrequencies~$\sigma$ as a
function of the azimuthal mode~$m$. For any~$m$, there is an infinite
set of eigenvalues. However, the corresponding eigenfunctions become
more and more oscillatory, implying larger and larger wave numbers. In
the numerical applications, we restrict our attention to the first few
that also correspond to the highest possible eigenvalues~$\sigma$.

\subsubsection{Newtonian vs general-relativistic disk}

We emphasize that the qualitative results presented in this paper are
independent of the nature of the spacetime, be it flat as in the
Newtonian gravitational potential or be it curved as in the Kerr
metric. We make this point clear by showing results according to both
spacetime structures. 

Let's start by recalling some facts about accretion disks orbiting in
the Kerr metric. When the inner edge of the accretion disk reaches a
few gravitational radii, general relativistic effects become
important.  The degeneracy between the three frequencies is lifted,
namely the orbital~$\Omega$, the radial epicyclic~$\kappa_r$, and the
vertical epicyclic~$\kappa_z$ frequencies. Their value depends not
only on the stellar mass~$M_*$ but also on the angular momentum of the
star~$a_*$.  Indeed we distinguish 3~characteristic frequencies in the
accretion disk around a Kerr black hole (or equivalently a rotating
neutron star):
\begin{itemize}
\item the orbital angular velocity:
  \begin{equation}
    \label{eq:Omega}
    \Omega(r,a_*)=\frac{1}{r^{3/2}+a_*} 
  \end{equation}
\item the radial epicyclic frequency:
  \begin{equation}
    \label{eq:KappaR}
    \kappa_r(r,a_*)=\Omega(r,a_*)\,\sqrt{1-\frac{6}{r} + 8\,\frac{a_*}{r^{3/2}}
      - 3\,\frac{a_*^2}{r^2} }  
  \end{equation}
\item the vertical epicyclic frequency:
  \begin{equation}
    \label{eq:KappaZ}
    \kappa_z(r,a_*)=\Omega(r,a_*)\,\sqrt{1 -4\,\frac{a_*}{r^{3/2}} +
      3\,\frac{a_*^2}{r^2}}
  \end{equation}
\end{itemize}
The parameter~$a_*$ corresponds to the angular momentum of the star
in geometrized units. For a neutron star, it is given by
$a_*=\frac{c\,I_*}{G\,M_*^2}\,\Omega_*$.

We have the following ordering~:
\begin{eqnarray}
  \Omega>\kappa_z>\kappa_r & {\rm for} & \;\; a_*>0 \\
  \kappa_z>\Omega>\kappa_r & {\rm for} & \;\; a_*<0
\end{eqnarray}

\subsubsection{Results}

The eigenvalues for the density waves are shown with decreasing
magnitude in Table~(\ref{tab:Eigenvalue}). This holds for a neutron
star with angular velocity $\nu_*=\Omega_*/2\pi=100~\mathrm{Hz}$. We
compared the Newtonian case with the Schwarzschild metric. The highest
speed pattern, given by~$\sigma/m$, never exceeds the orbital
frequency at the ISCO. Note that in the~$m=1$ case, the inner Lindblad
resonance does not exist for the Newtonian potential, and only a few
free wave solutions can be found for the Schwarzschild geometry.

\begin{table}[htbp]
  \caption{The first eight largest eigenvalues~$\sigma$ for 
    the free wave solutions of Eq.~(\ref{eq:AppPsiSimp}).
    Values are normalized to the frequency of the ISCO,~$\Omega_{isco}=6^{-3/2}$.
    Results are given for 3~azimuthal modes~$m=1,2,5$ for 
    the Newtonian, as well as for the Schwarzschild gravitational field.
    Where no inner Lindblad resonance exist, this is represented by~$-$.}
  \label{tab:Eigenvalue}
  \centering
  \begin{tabular}{c c c | c c c}
    \hline
    \hline
    \multicolumn{6}{c}{Eigenvalues $\sigma/\Omega_{isco}$} \\
    \hline
    \multicolumn{3}{c|}{Newtonian} & \multicolumn{3}{c}{Schwarzschild} \\
    \hline
    $m=1$ & $m=2$ & $m=5$ & $m=1$ & $m=2$ & $m=5$ \\
    \hline
    \hline
    -----  &  0.745  &  3.286  &  0.3727 &  1.139  &  3.643 \\
    -----  &  0.484  &  2.564  &  0.126  &  0.704  &  2.802 \\
    -----  &  0.358  &  2.155  &  0.0528 &  0.504  &  2.330 \\
    -----  &  0.278  &  1.870  &  0.0217 &  0.382  &  2.008 \\
    -----  &  0.220  &  1.652  &  0.010  &  0.298  &  1.765 \\
    -----  &  0.177  &  1.476  &  -----  &  0.236  &  1.571 \\
    -----  &  0.143  &  1.330  &  -----  &  0.189  &  1.411 \\
    -----  &  0.116  &  1.206  &  -----  &  0.152  &  1.276 \\
    \hline
  \end{tabular}
\end{table}

\begin{figure}[htbp]
  \centering
  \begin{tabular}{cc}
    \includegraphics[scale=0.5]{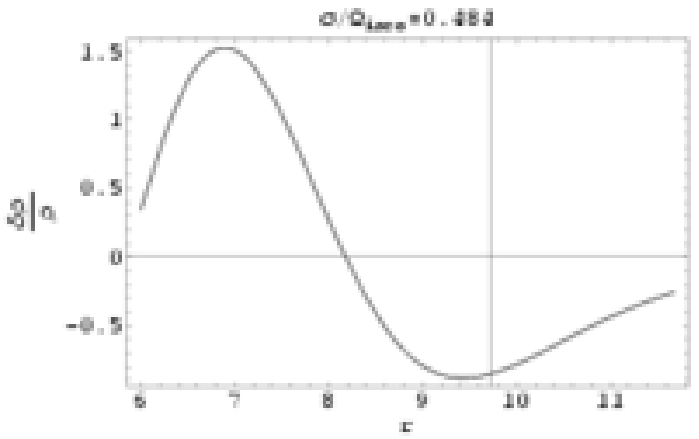} &
    \includegraphics[scale=0.5]{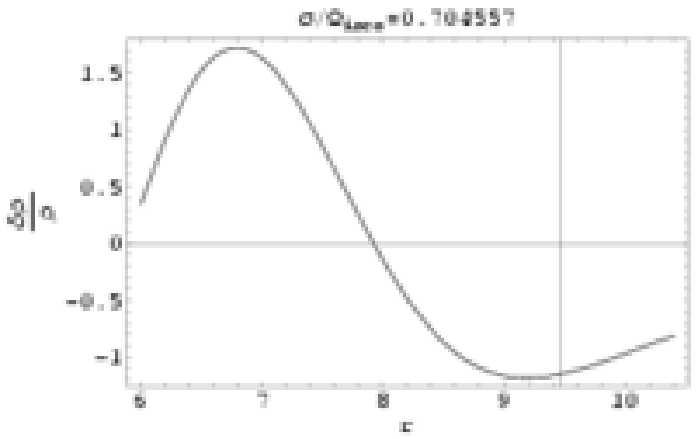} \\
    \includegraphics[scale=0.5]{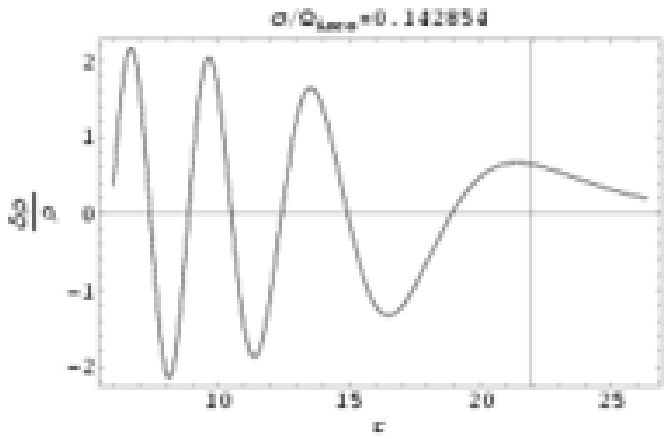} &
    \includegraphics[scale=0.5]{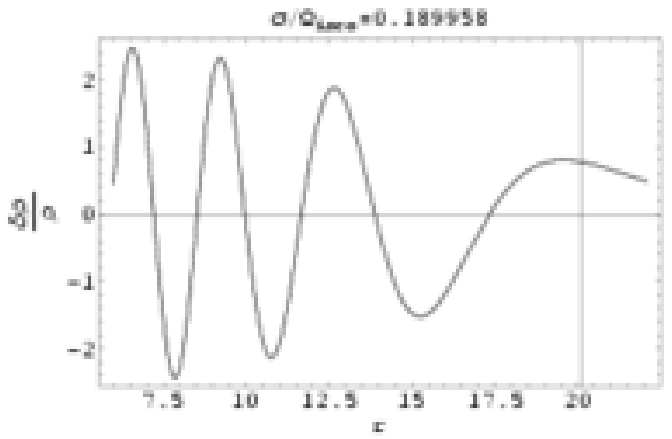}
  \end{tabular}
  \caption{Density wave perturbation in the disk caused by 
    the free wave propagation for the azimuthal mode~$m=2$.  Some
    examples are shown for different eigenvalues and for Newtonian
    geometry, on the left, as well as for Schwarzschild geometry, on
    the right.  The vertical bar indicates the location of the inner
    Lindblad resonance. The normalization of the eigenfunctions is
    arbitrary.}
  \label{fig:FnPropre}
\end{figure}

Some examples of the corresponding eigenfunctions for density waves
are shown in Fig.~\ref{fig:FnPropre} with arbitrary normalization.
Each of them possesses its own inner Lindblad radius depending on the
eigenvalue. They are distinguished by the number of radial nodes they
possess, starting from no node at all, which corresponds to the
highest speed pattern.

Because the pattern speed of the first spiral waves is close to the
ISCO, we suggest that these free wave solutions can be associated with
the highest kHz QPO. More precisely, there exists one eigenfunction
which possesses no root, namely the one with the highest eigenvalue.
Eigenfunctions with several nodes in the radial distribution will emit
no significant radiation since on average the fluctuations will cancel
out. Therefore, we expect to see only density fluctuations with no or
very few radial nodes. The most interesting candidate is the one with
no node, a situation that also corresponds to the highest eigenvalue.

Moreover, while propagating in the perturbed rotating gravitational
field of the star, this spiral wave experiences a sinusoidally varying
gravity at a rate~$\Omega_p-\Omega_*$, where $\Omega_p$ is the speed
pattern of the wave. This induces a modulation in the shape of the
wave which will be reflected in the power spectrum density as a kind of
beat phenomenon.  As a result, in our model, we associate the highest
kHz-QPO to $\nu_2=\Omega_p$ and the lowest kHz QPO to the beat
frequency~$\nu_1=\Omega_p-\Omega_*$. The peak separation is then
$\Delta\nu=\Omega_*$.

In a real accretion disk, the precise location of the inner edge does
not necessarily reach the ISCO.  The magnetic field or the radiation
pressure can prevent the disk from doing so. Indeed, the boundary of
the inner part of the disk can fluctuate due to the varying accretion
rate. For instance when the accretion process is enhanced, the inner
edge moves closer to the ISCO. As a result, the highest eigenvalue of
the free waves also increases and therefore~$\nu_2$ shifts to higher
frequencies as well as~$\nu_1$.

\begin{figure}[htbp]
  \centering
  \includegraphics[draft=false,scale=1]{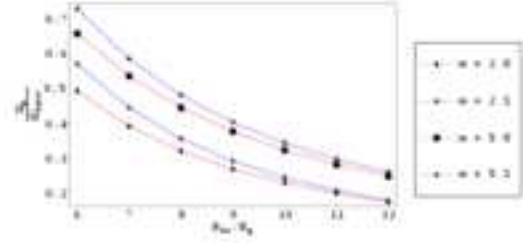}
  \caption{Variation of the highest eigenvalue, corresponding to 
    the eigenfunction having no node, as a function of the location of
    the inner edge of the disk~$R_{in}$. There is a monotonic increase
    as the disk approaches the ISCO at~$R_{in}=6R_g$. Results are
    shown for the~$m=2,5$ modes in the Newtonian (N) and Schwarzschild
    (S) spacetime, red and blue curves respectively. The gravitational
    radius is defined by~$R_g=G\,M_*/c^2$.}
  \label{fig:SigvsR1}
\end{figure}

Examples of this mechanism are shown in Fig.~\ref{fig:SigvsR1}.  We
estimate the highest eigenvalue for the modes~$m=2,5$ in the
Newtonian, as well as in the Schwarzschild gravitational potential.
The pattern speed of the wave solution~$\Omega_p$, normalized to the
frequency of the ISCO,~$\Omega_{isco}$, is shown as a function of the
location of the inner boundary of the accretion disk~$R_{in}$. In each
case, we observe an increase in the eigenvalue when the disk's inner
edge moves closer to the compact object, corresponding to an increased
accretion rate. When the ISCO is reached, the boundary condition does
not change anymore and the eigenvalue saturates to its final value.
Thus the kHz-QPO frequencies associated with the speed pattern of the
wave saturate.  This has been observed in some LMXBs, as reported for
instance in a paper by Zhang et al.~(\cite{Zhang1998}).

\subsubsection{Non-wavelike perturbation}
\label{sec:Nonwave}

\begin{figure}[htbp]
  \centering
  \includegraphics[draft=false,scale=1]{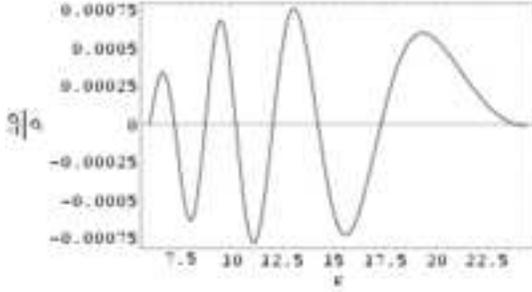}  
  \caption{Non-wavelike disturbance in the inner part of the accretion disk
    due to the asymmetric component of the magnetic field. The
    potential is Newtonian and~$m=2$. The numerical values used are
    described in Sec.\ref{sec:Simulation}.}
  \label{fig:NonWave}
\end{figure}

We now solve the full inhomogeneous Eq.~(\ref{eq:XiRSchroed}) in order
to find the solution corresponding to the non-wavelike perturbation.
In this case, the eigenvalue is known:~$\sigma = m\,\Omega_*$.  We
have to solve the second order ordinary differential equation
for~$\psi$ with the appropriate boundary conditions
Eq.~(\ref{eq:Boundary}). Following the results given in
Appendix~\ref{app:AppSchroeApprox}, the most general solution is:
\begin{eqnarray}
  \label{eq:PsiPart}
  \psi_r(r) = C_1 \, \psi_1(r) + C_2 \, \psi_2(r) +
  \pi \, \mathrm{sign}(\omega_1'(r)) \, \times & & \nonumber \\
  \int_{r_L}^r \left( \psi_1(r) \, \psi_2(s) - \psi_1(s) \, \psi_2(r) \right) F(s) \, ds. & &
\end{eqnarray}
The constant~$C_2$ is chosen such that the solution remains bounded
for~$r\gg r_L$:
\begin{equation}
  \label{eq:C2}
  C_2 = \lim_{r\to\infty} \pi \, \mathrm{sign}(\omega_1') \, \int_{r_L}^r
  \psi_1(s) \, F(s) \, ds.
\end{equation}
This integral is convergent because the function~$\psi_1$ is
exponentially decreasing with the radius. We can directly integrate
this equation to obtain the shape of the density perturbation between
the inner edge of the accretion disk and the inner Lindblad resonance.
The results for a particular choice of the density and magnetic
profiles (discussed in Sec.\ref{sec:Simulation}) are shown in
Fig.~\ref{fig:NonWave}.  This can be used later to check the numerical
algorithm. It also gives a good estimate of the amplitude of these
fluctuations because it is directly related to the strength of the
stellar magnetic field asymmetry. The maximum density fluctuation is
about~0.1\%. The oscillations become evanescent upon reaching the
inner Lindblad resonance.

\subsubsection{Counterrotating disk}
\label{sec:ContreRot}

\begin{figure}[htbp]
  \centering
  \includegraphics[draft=false,scale=1]{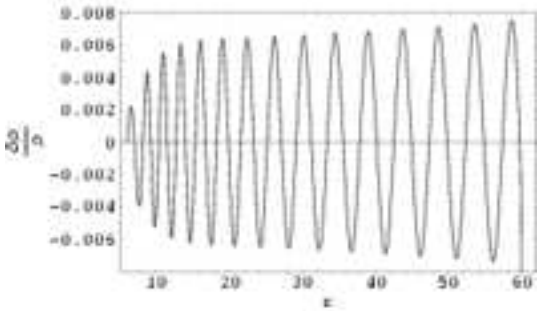}  
  \caption{Non-wavelike disturbance in the inner part of the accretion disk
    due to the asymmetric component of the magnetic field. The
    potential is Newtonian,~$m=2$ and the disk is counterrotating.}
  \label{fig:NonWaveContre}
\end{figure}

The counterrotating disk can be solved in exactly the same way by
integrating Eq.~(\ref{eq:XiRSchroed}) to find the solution
corresponding to the non-wavelike perturbation. The amplitude of the
oscillations is shown in Fig.~\ref{fig:NonWaveContre}. The Lindblad
resonances are no longer in the computational domain. An oscillatory
structure appears in the whole disk. The maximum amplitude is
about~1\% of the average density. This can later be compared with the
numerical simulations.

\section{SIMPLIFIED ANALYSIS}

To get more insight into the nature of the resonances, we now focus on
the displacement of the disk either in the equatorial plane or
perpendicular to it. This means that we neglect the coupling between
the oscillations occurring in perpendicular directions. This
simplification will help us to understand the origin of the
instability and to derive some resonance conditions.

To study the motion in the vertical direction, setting~$(\xi_r,
\xi_\varphi) = 0$ we find:
\begin{eqnarray}
  \label{eq:PDEXiz}
   \rho \, \frac{D^2\xi_z}{Dt^2} - \frac{\partial}{\partial z} \left[ 
    \left( \rho \, c_{mar}^2 + \frac{B_r\,\delta B_*^r}{\mu_0} \right)
    \frac{\partial\xi_z}{\partial z} \right] + 
  \rho \, \kappa_z^2 \, \xi_z & = & \nonumber \\
  \frac{\partial}{\partial z} \left( \frac{{\delta B_*^r}^2 + 
      {\delta B_*^\varphi}^2}{2\,\mu_0} +
    \frac{B_r\,\delta B_*^r}{\mu_0} \right) + \frac{\partial}{\partial z}
  \left( \frac{\delta B_*^r }{\mu_0} \, 
    \frac{\partial B_r}{\partial z} \, \xi_z \right)
\end{eqnarray}
See Appendix~\ref{app:Eigenvalue} for the details.  The Alfven
velocity associated with the radial magnetic field and the radial fast
magneto-acoustic wave speed are, respectively:
\begin{eqnarray}
  c_{ar}^2 & = & \frac{B_r^2}{\mu_0\,\rho} \\
  c_{mar}^2 & = & c_s^2 + c_{ar}^2
\end{eqnarray}

The same can be done for the orbital motion by setting~$\xi_z=0$,
which leads to a similar expression~:
\begin{eqnarray}
  \label{eq:PDEXir}
  \rho \, \frac{D^2\xi_r}{Dt^2} - \frac{\partial}{\partial r} \left[ 
    \left( \rho \, c_{maz}^2 + \frac{B_z\,\delta B_*^z}{\mu_0} \right)
    \, \frac{1}{r} \, \frac{\partial}{\partial r} \left( r \, \xi_r \right) \right] + 
  \rho \, \kappa_r^2 \, \xi_r
  & = & \nonumber \\
  - \frac{\partial}{\partial r} \left( \frac{{\delta B_*^z}^2}{2\,\mu_0} +
    \frac{B_z\,\delta B_*^z}{\mu_0} \right) + \frac{\partial}{\partial r}
  \left( \frac{\delta B_*^z}{\mu_0} \, 
    \frac{\partial B_z}{\partial r} \, \xi_r \right)
\end{eqnarray}
Eq.~(\ref{eq:PDEXiz}) and~(\ref{eq:PDEXir}) look very similar, the
discrepancy coming only from the difference between the planar and the
cylindrical geometry (terms containing~$r$).  In the two first terms
of Eq.~(\ref{eq:PDEXiz}) and~(\ref{eq:PDEXir}) we recognize an MHD
wave propagation in a tube of spatially varying but time independent
cross section~(Morse \& Feshbach~\cite{Morse1953}). The first and
third terms put together form an harmonic oscillator at the epicyclic
frequency. So the three first terms are a generalization of the
Klein-Gordon equation and do not give rise to any kind of instability.
The interesting parts are those containing the perturbation in the
magnetic field~$\delta\vec{B}_*$.

Further, we neglect the magneto-acoustic wave propagation because of
the weakly magnetized thin disk approximation, and also assume a
homogeneous medium for which the vertical Lagrangian displacement and
density are independent of the altitude~$z$ in the first case and
where the density and radial Lagrangian displacement do not depend on
the radius~$r$ in the second case. Note also that terms like~${\delta
  B_*^i}^2$ are second order with respect to the perturbation and
should be neglected. These simplifications make the equations much
more easier to manipulate without losing the most interesting physical
aspect contained in the resonance terms. We can then simplified
Eq.~(\ref{eq:PDEXiz}) and Eq.~(\ref{eq:PDEXir}) to obtain~:
\begin{eqnarray}
  \label{eq:PDEXiRZ}
  \frac{D^2\xi_z}{Dt^2} + \left[ \kappa_z^2 -  \frac{\partial}{\partial z}
    \left( \frac{\delta B_*^r }{\mu_0\,\rho} \, 
      \frac{\partial B_r}{\partial z} \right) \right] \, \xi_z & = & 
  \frac{\partial}{\partial z} \left( 
    \frac{B_r\,\delta B_*^r}{\mu_0\,\rho} \right) \\
  \frac{D^2\xi_r}{Dt^2} + \left[ \kappa_r^2 - \frac{\partial}{\partial r}
    \left( \frac{\delta B_*^z}{\mu_0\,\rho} \, \frac{\partial B_z}{\partial r}
    \right) \right] \, \xi_r & = &
  - \frac{\partial}{\partial r} \left( \frac{B_z\,\delta B_*^z}{\mu_0\,\rho} \right).
\nonumber \\
 & & 
\end{eqnarray}
Recall that the magnetic perturbation is dragged with the star's
rotation.  Therefore, when looking at a given fixed point in the
accretion disk, the coefficients of the equations containing~$\delta
B_*^i$ will vary periodically.  Thus, we recognize a kind of Hill
equation corresponding to an oscillator with periodically time-varying
eigenfrequency and subject to a periodic driven force. It is well
known that this type of equation shows what is called a parametric
resonance.  Moreover, due to the rotation of the star, this
perturbation will vary sinusoidally in time. In fact, in the frame
locally corotating with the disk, (convective derivative), the
modulation follows a harmonic temporal behavior such as~:
\begin{eqnarray}
  \cos(\varphi - (\Omega-\Omega_*)\,t) & , & 
  \sin(\varphi - (\Omega-\Omega_*)\,t).
\end{eqnarray}
The Hill equation specializes to Mathieu's equation for which we can
derive the resonance conditions analytically. The solutions to
Mathieu's equation written in the form~$y''(t) + \omega_0^2 ( 1 + h \,
\cos \gamma\,t) \, y(t) = 0$ will indeed become unbounded if~$\gamma =
2 \frac{\omega_0}{n}$ where~$n$ is an integer, (Landau \&
Lifshitz~\cite{Landau1982}).

The equation satisfied by each component of the Lagrangian
displacement when looking at a given azimuth~$\varphi_0$ is therefore
in the form~:
\begin{eqnarray}
  \label{eq:Typ}
  \xi''(t) + \left[ \kappa^2 + h \, 
    \left\{ 
      \begin{array}{c}
        \sin \\
        \cos
      \end{array}
    \right\}
    (\varphi_0 - ( \Omega - \Omega_* ) \, t ) 
  \right] \, \xi(t) = & & \nonumber \\
  f \, 
  \left\{ 
    \begin{array}{c}
      \sin \\
      \cos
    \end{array}
  \right\}
  (\varphi_0 - ( \Omega - \Omega_* ) \, t ),
\end{eqnarray}
where $f(\delta B_*^i)$ and~$h(\delta B_*^i)$ are two constants
depending on the strength of the stellar magnetic perturbation.

From this analysis we expect three kinds of resonances corresponding
to, (see Appendix~\ref{App:Resonances} for a detailed analysis)~:
\begin{itemize}
\item a {\it corotation resonance} at the radius where the angular
  velocity of the disk equals the rotation speed of the star. This is
  only possible for prograde disks. The resonance condition to
  determine the corotating radius is simply
  \begin{equation}
    \label{eq:ResCorot}
    \Omega = \Omega_* .
  \end{equation}
\item an {\it inner and an outer Lindblad resonance} at the radii
  where the radial or vertical epicyclic frequency equals the
  frequency of the dipolar magnetic perturbation as measured in the
  locally comoving frame. Due to the~$m=1$ structure of this
  perturbation we find the resonance condition to be
  \begin{equation}
    \label{eq:ResForcage}
    |\Omega-\Omega_*| = \kappa_{r/z}.
  \end{equation}
\item a {\it parametric resonance} related to the time-varying
  epicyclic frequency, (Hill equation). The rotation of the star
  induces a sinusoidally time varying epicyclic frequency leading to
  the well-known Mathieu equation. The resonance condition is then
  \begin{equation}
    \label{eq:ResPara}
    |\Omega-\Omega_*| = 2 \, \frac{\kappa_{r/z}}{n}.
  \end{equation}
\end{itemize}
Note that the driven resonance occurs at the same radius as does the
parametric resonance for~$n=2$. However, their growth rates are
different, because forcing causes a linear growth in time of the
amplitude while parametric resonance causes an exponential growth.

More generally, for a perturbation of arbitrary azimuthal mode~$m$,
the resonance conditions are given by~:
\begin{itemize}
\item for the inner and outer Lindblad resonances~: $ m\,| \Omega -
  \Omega_* | = \kappa_{r/z} $
\item for the parametric resonance~: $ m\,|\Omega - \Omega_*| =
  2\,\frac{\kappa_{r/z}}{n} $.
\end{itemize}
The conclusion of this study is that the locations of the resonances
are determined solely by the rotation rate of the neutron star.
Fluctuations of the accretion rate are not included in this picture.
However, we will show by a more detailed linear analysis that there
are some free-wave solutions for the density perturbation for which
the eigenvalue can be correlated to the accretion rate in accordance
with observations, see Sec.~\ref{sec:ThinDisk}.

\subsection{Results}

\subsubsection{Newtonian disk}

From the resonance conditions derived previously,
Eq.~(\ref{eq:ResCorot})-(\ref{eq:ResPara}), we can estimate the radii
where each of these resonances occurs. Beginning with the Newtonian
potential, it is well known that the angular velocity, the radial and
epicyclic frequencies are all equal. For a weakly magnetized thin
accretion disk, this statement remains true so that~$\Omega \approx
\kappa_r \approx \kappa_z$.  Distinguishing between the two signs of
the absolute value, we get the following rotation rate for the
parametric condition~(\ref{eq:ResPara}):
\begin{eqnarray}
  \frac{\Omega}{\Omega_*} & = & \frac{n}{n+2} = 1/3, 1/2, 3/5, 2/3, 5/7, ... \\
  \frac{\Omega}{\Omega_*} & = & \frac{n}{n-2} = -1 , - , 3, 2, 5/3, ...
\end{eqnarray}
As a consequence, each resonance will occur in the frequency
range~$[\Omega_*/3, 3\,\Omega_*]$. Assuming that the QPOs are related
to the orbital motion and taking into account that the spatial
structures of the resonances are formed by~$m=1$ patterns, we expect
to see QPOs in the same range~$[\nu_*/3, 3\,\nu_*]~\mathrm{Hz}$ where
we have introduced the spin frequency by~$\nu_* = \Omega_*/2\pi$.  For
instance, for a~$300~\mathrm{Hz}$ spinning neutron star, we can expect
QPOs between~$[100,900]~\mathrm{Hz}$.  However, observations have
shown that QPOs with frequency up to 1300~Hz exist for this kind of
system. We will see in the next subsection that this difficulty can be
solved by introducing the general relativistic effects of the
innermost stable circular orbit. Table~\ref{tab:ResPara} gives an
example of the expected orbital frequencies where the resonances are
expected to occur for a neutron star spinning at 300 and 600~Hz.

\begin{table}
  \caption{Value of the orbital frequencies at the parametric
    resonance radii for the first five orders~$n$ in the case of a
    Newtonian gravitational potential.  The results are given for
    a~$1.4\,\mathrm{M_\odot}$ neutron star rotating  at~$300$
    and~$600~\mathrm{Hz}$ respectively. The value to the left of the symbol~/ corresponds
    to the absolute value sign taken to be~$-$ and on the right to
    be~$+$. The $---$ sign indicates that no resonance condition can be satisfied. }
  \label{tab:ResPara}
  \centering  
  \begin{tabular}{c c c}
    \hline
    \hline
    rank n & \multicolumn{2}{c}{Orbital frequency $\nu(r,a_*)~\mathrm{(Hz)}$} \\
    \hline
    & $\nu_*=600~\mathrm{Hz}$ & $\nu_*=300~\mathrm{Hz}$ \\
    \hline
    \hline
    1 & -600 / 200 & -300 / 100 \\
    2 & ---- / 300 & ---- / 150 \\ 
    3 & 1800 / 360 &  900 / 180 \\
    4 & 1200 / 400 &  600 / 200 \\
    5 & 1000 / 429 &  500 / 214 \\
    \hline
  \end{tabular}
\end{table}

\subsubsection{General relativistic disk}

The parametric resonance condition Eq.~(\ref{eq:ResPara}), split into
the two cases depending on the sign of the absolute value, becomes:
\begin{eqnarray}
\label{eq:ResParaGR1}
\Omega(r,a_*) - 2 \, \frac{\kappa_{r/z}(r,a_*)}{n} & = & \Omega_* \\
\label{eq:ResParaGR2}
\Omega(r,a_*) + 2 \, \frac{\kappa_{r/z}(r,a_*)}{n} & = & \Omega_*
\end{eqnarray}

For a typical neutron star, we choose:
\begin{itemize}
\item mass~$M_*=1.4\,\mathrm{M_\odot}$~;
\item angular velocity~$\nu_*=\Omega_*/2\pi=300-600~\mathrm{Hz}$~;
\item moment of inertia~$I_*=10^{38}\;\mathrm{kg\,m^2}$~;
\item angular momentum~$a_*=\frac{c\,I_*}{G\,M_*^2}\,\Omega_*$.
\end{itemize}
The angular momentum is then given by~$a_*=5.79*10^{-5}\,\Omega_*$.
For a given angular momentum~$a_*$, we have to solve
equations~(\ref{eq:ResParaGR1}),(\ref{eq:ResParaGR2}).  Solving them
for the radius~$r$ and then deducing the orbital frequency at this
radius, we get the results shown in Table~\ref{tab:ResParaGR}. For the
spin rate of the star we find~$a_*=0.109-0.218$ and so the vertical
epicyclic frequency is close to the orbital
one~$\kappa_z\approx\Omega$. Thus for the vertical resonance, we are
still close to the Newtonian case explored in the previous section.

\begin{table}
  \caption{Value of the orbital frequencies at the parametric
    resonance radii for the first five orders~$n$ in the general
    relativistic case.  The results are given for a~$1.4\,\mathrm{M_\odot}$
    neutron star rotating  at~$300$ and~$600~\mathrm{Hz}$ respectively. The
    value to the left of the symbol~/ corresponds to the absolute
    value sign taken to be~$-$ and on the right to be~$+$.}
  \label{tab:ResParaGR}
  \centering
  \begin{tabular}{c | c c | c c}
    \hline
    \hline
    rank n & \multicolumn{4}{c}{Orbital frequency $\nu(r,a_*)$ (Hz)} \\
    \hline
    & \multicolumn{2}{c|}{Vertical} & \multicolumn{2}{c}{Radial} \\
    & $\nu_*=600$~Hz & $\nu_*=300$~Hz & $\nu_*=600$~Hz & $\nu_*=300$~Hz \\
    \hline
    \hline
    1 & ---- / 200 & --- / 100 & 1596 / 220 & 1332 / 106 \\
    2 & ---- / 301 & --- / 150 & 1196 / 330 &  800 / 159 \\ 
    3 & 1695 / 361 & 885 / 180 &  980 / 392 &  573 / 190 \\
    4 & 1175 / 401 & 597 / 200 &  870 / 432 &  480 / 210 \\
    5 &  988 / 430 & 498 / 214 &  808 / 459 &  433 / 223 \\
    \hline
  \end{tabular}
\end{table}

We have now overcome the problem faced in the Newtonian approximation.
Indeed, even in the case of a $300$~Hz spinning neutron star, we can
expect frequencies as high as $1332$~Hz for the~$n=1$ rank, in
accordance with observations. The general-relativistic effect caused
by the presence of the ISCO is therefore a crucial characteristic that
makes a clear distinction with the Newtonian potential. The ISCO has
to be implemented in any realistic QPO model in order to get a good
quantitative agreement with observations.  Therefore, a fully
general-relativistic simulation code should be developed in order to
take this ISCO into account in a self-consistent manner. However, we
point out that general relativistic effects are not directly
responsible for the observed QPOs as claimed in other models~(Lee et
al~\cite{Lee2004}, Stella et al.~\cite{Stella1999b}).  They merely
lead to quantitative improvements compared to the flat spacetime
results.

\section{NUMERICAL SIMULATIONS}
\label{sec:Simulation}

The linear analysis described above is not the end of the story. It
was shown that the disk becomes unstable at some preferred radii where
the resonance conditions are satisfied.  But this analysis does not
tell us what will happen to these resonances on a long time scale.
Moreover, as in the hydrodynamical case, the non-linearities in the
time evolution of the disk will give rise to a rich variety of
physical processes.  However, in this simulation, in order to check
our numerical algorithm, we start with runs for which the perturbation
remains weak, in order to keep a regime close to the linear one
described in the previous section. The full non-linear effects arising
from a strong magnetic field perturbation will be treated in a
forthcoming paper.

To achieve this goal, we performed 2D numerical simulations by solving
the magnetohydrodynamical equations for the accretion disk
Eq.~(\ref{eq:DiscMHDrho})-(\ref{eq:DiscMHDB}). This is done by a
pseudo-spectral method for a simplified version of the 2D MHD
accretion disk.  Indeed, an exact treatment of the magnetic field
induced by the motions in the disk would involve solution of an
elliptic partial differential equation related to Ampere's law.  We
don't want to go into such refinement in this paper but only emphasize
the role of a rotating asymmetric magnetic field. Nevertheless the
physics at hand remains the same.

\subsection{Boundary and initial conditions}

We expand every quantity in Fourier series in the azimuthal direction
and in Tchebyshev series in the radial direction, whereas time
integration is performed by an explicit fourth-order Runge-Kutta
scheme.  In this work we are not interested in the accretion process
itself but focus only on the oscillations occurring in this system
around its stationary state.  We can then impose fixed boundary
conditions in the radial direction by setting the radial velocity
equal to zero at the two boundaries.  Furthermore, to avoid spurious
reflections at the edges of the accretion disk, we force the radial
velocity to decrease to zero in a narrow layer around these
boundaries. We also require the density and the azimuthal velocity to
reach their equilibrium. This acts as an ad hoc absorption process
damping out the strong oscillations which could arise in these
regions. The non-reflecting boundaries conditions are therefore
expressed as:
\begin{eqnarray}
  \label{eq:Boundaries}
  \rho(r,\varphi) & = & f(r) \, ( \tilde{\rho}(r,\varphi) - \rho^{eq}(r,\varphi) ) + 
  \rho^{eq}(r,\varphi) \\
  v_r(r,\varphi) & = & f(r) \, \tilde{v}_r(r,\varphi) \\
  v_\varphi(r,\varphi) & = & f(r) \, ( \tilde{v}_\varphi(r,\varphi) - 
  v_\varphi^{eq}(r,\varphi) ) + v_\varphi^{eq}(r,\varphi) \\
  B_z(r,\varphi) & = & f(r) \, ( \tilde{B}_z(r,\varphi) - 
  B_z^{eq}(r,\varphi) ) + B_z^{eq}(r,\varphi)
\end{eqnarray}
The quantities denoted by an~$\tilde{X}$ subscript correspond to the
physical values before applying the non-reflecting boundary
conditions.  Those values at equilibrium are denoted by~$X^{eq}$.
Function~$f$ acts as a filter, as its value is one everywhere except
close to the inner and outer boundaries where it decreases to zero in
a narrow transition layer.

We adopted the geometry of a cylinder of gas threaded by a vertical
magnetic field and a radially directed gravitational field.  The
problem is invariant under translation in the direction of the
magnetic field lines, which is chosen to be the $z$~direction,
~$\partial/\partial z=0$. Motions of the gas are only allowed in the
equatorial plane~$(r,\varphi)$, $v_z=0$; therefore, we suppress the
possibility of a warped disk and a possible precession of its orbital
plane, leaving this to future work.  The ideal MHD
equations~(\ref{eq:DiscMHDrho}) and (\ref{eq:DiscMHDv}) then simplify
to~:
\begin{eqnarray}
  \label{eq:MHD2D}
  \frac{\partial \rho}{\partial t} + \frac{1}{r}\,\frac{\partial}{\partial r}
  (r\,\rho\,v_r) + \frac{1}{r} \, \frac{\partial}{\partial \varphi} (\rho\,v_\varphi) & = & 0 \\
  \frac{\partial v_r}{\partial t} + v_r \, \frac{\partial v_r}{\partial r}
  + \frac{v_\varphi}{r} \, \frac{\partial v_r}{\partial\varphi} - \frac{v_\varphi^2}{r} 
  & = & g_r - \frac{1}{\rho} \, \frac{\partial p}{\partial r} + 
  \frac{j_\varphi \, B_z}{\rho} \nonumber \\
 & & \\
  \frac{\partial v_\varphi}{\partial t} + v_r \, \frac{\partial v_\varphi}{\partial r}
  + \frac{v_\varphi}{r} \, \frac{\partial v_\varphi}{\partial\varphi} 
  + \frac{v_r \, v_\varphi}{r} 
  & = & g_\varphi - \frac{1}{\rho\,r} \, \frac{\partial p}{\partial \varphi} -
  \frac{j_r \, B_z}{\rho} \nonumber \\
 & &
\end{eqnarray}
Moreover, we adopt a polytropic equation of state for the plasma which
is:
\begin{equation}
  \label{eq:GasPoly}
  p = K \, \rho^\gamma.
\end{equation}
The induction equation~(\ref{eq:DiscMHDBind}) in this particular
geometry transforms into an advection equation for the vertical
component~$B_z$, similar to the mass conservation. This means that the
magnetic flux is conserved during time evolution:
\begin{eqnarray}
  \frac{\partial B_z}{\partial t} + \frac{1}{r}\,\frac{\partial}{\partial r}
  (r\,B_z\,v_r) + \frac{1}{r} \, \frac{\partial}{\partial \varphi} (B_z\,v_\varphi) & = & 0 \\
  \frac{\partial B_r}{\partial t} & = & 0 \\
  \frac{\partial B_\varphi}{\partial t} & = & 0 .
\end{eqnarray}
Inspecting these expressions, we conclude that there is no radial or
azimuthal magnetic field generated during the evolution of the disk.
Because at the initial conditions~$B_r=B_\varphi=0$, these components
will remain equal to zero during the whole simulation. Consequently
the divergence free condition of~$\vec{B}$ is automatically satisfied
by:
\begin{equation}
  \label{eq:DivB}
  \div \vec{B} = \frac{\partial B_z}{\partial z} = 0.
\end{equation}
Finally, applying Ampere's law, Eq.~(\ref{eq:DiscMHDB}), we get the
current density in the disk by:
\begin{eqnarray}
  j_r & = & \frac{1}{\mu_0 \, r} \, \frac{\partial B_z}{\partial\varphi} \\
  j_\varphi & = & - \frac{1}{\mu_0} \, \frac{\partial B_z}{\partial r} \\
  j_z & = & 0 
\end{eqnarray}
We expand every quantity in a Fourier series in the azimuthal
direction and in a Tchebyshev series in the radial direction.
Moreover, to avoid the spectral aliasing that arises from the
non-linearities of the equations, we use the 2/3-rule to completely
wash out the highest frequencies~(Gottlieb and
Orszag~\cite{Gottlieb1977}, Canuto et al.~\cite{Canuto1990})
(desaliasing process).

We use geometrized units in which~$G=c=1$. Distance is measured in
units of the gravitational radius given by~$R_g=G\,M_*/c^2$.
Moreover, the simulations are done for a star with~$M_*=1$, so that in
the new units we have~$R_g=1$.  The star is assumed to be perfectly
spherical with the three main axes all equal~$R_x=R_y=R_z$, and the
perturbation comes from the magnetic field alone. The adiabatic
constant is equal to~$\gamma=5/3$ and typical resolution
is~$256\times32$.

We checked that this resolution is sufficient by running a simulation
with twice as many points in both directions, namely~$512\times64$. We
found no changes in the solutions. Indeed inspecting the
Fourier-Tchebyshev coefficients, only half of the radial coefficients
and a few of the azimuthal coefficients in the 2D expansions are
significantly different from zero. Choosing a higher resolution will
not lead to any improvement in the solution but only to excessive
computation time.

The initial density profile is prescribed as
\begin{equation}
  \label{eq:DensiteInit}
  \rho_0(r) = \frac{10^{-6}}{r}.
\end{equation}
Initially the contribution to the magnetic field is given by the star
and the disk as follows
\begin{eqnarray}
  \label{eq:Binit}
  B_z^* & = & - 5 \, \frac{10^{-3}}{r^3} \\
  B_z^d & = & \frac{10^{-3}}{r^3},
\end{eqnarray}
while the stellar magnetic perturbation is given by
\begin{eqnarray}
  \label{eq:deltaBinit}
  \delta B_z^* & = & \delta b \, B_z^* \, \cos( m\,(\varphi - \Omega_* \, t )) \\
  \delta b & = & 0.001   .
\end{eqnarray}
The $r$-dependence of the magnetic perturbation
Eq.~(\ref{eq:deltaBinit}) is chosen to mimic the rotation of a
misaligned rotator rotating at a rate~$\Omega_*$ with a typical decay
as~$r^{-3}$. All other components are equal to zero.  The azimuthal
mode of the magnetic field perturbation~$\delta B_z^*$ can be chosen
arbitrarily. We use the case~$m=2$ to directly compare with the
hydrodynamical simulations. However, the~$m=1$ case is better adapted
to the true dipolar structure of the stellar magnetic field.  We will
give some examples for this situation, too.

\begin{figure}[htbp]
  \centering
  \begin{tabular}{cc}
    \includegraphics[draft=false,scale=0.5]{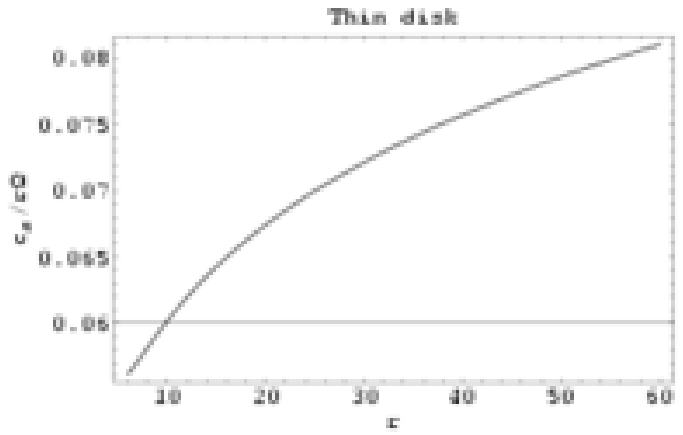} &
    \includegraphics[draft=false,scale=0.5]{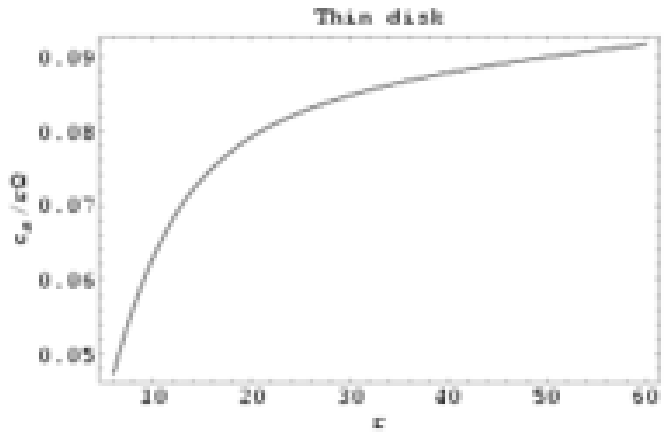}
  \end{tabular}
  \caption{Thin disk approximation in which the ratio~$H/R$ 
    never exceeds~$0.1$, on the left for the Newtonian potential and on
    the right for the Schwarzschild geometry.  }
  \label{fig:HsR}
\end{figure}

\begin{figure}[htbp]
  \centering
  \begin{tabular}{cc}
    \includegraphics[draft=false,scale=0.5]{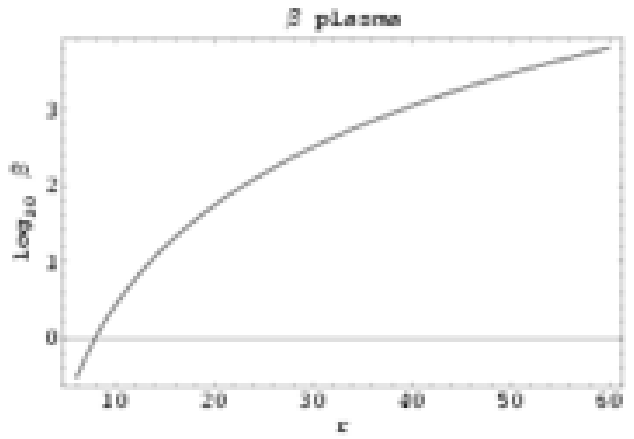} &
    \includegraphics[draft=false,scale=0.5]{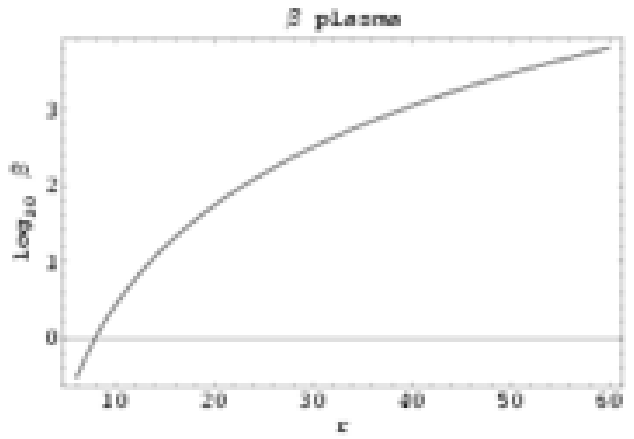}
  \end{tabular}
  \caption{Plasma~$\beta$ parameter for the weakly magnetized disk 
    plotted on a logarithmic scale.  The magnetic field dominates only
    very close to the inner edge, on the left for the Newtonian potential and on
    the right for the Schwarzschild geometry.}
  \label{fig:BetaPlasma}
\end{figure}

With these prescribed initial conditions, the thin disk approximation
is fairly good. Indeed the disk ratio~$H/R$, $H$ typical height and
$R$ radius of the disk, plotted for the Newtonian and Schwarzschild
geometry in Fig.~(\ref{fig:HsR}), never exceed~$0.1$ in any run.
Moreover, the disk is threaded by a weak magnetic field except very
close to the inner edge. The plasma~$\beta$ parameter is shown in
Fig.~\ref{fig:BetaPlasma} and remains very high.

Before the time~$t=0$, the disk stays in its axisymmetric equilibrium
state and possesses only an azimuthal motion. At~$t=0$, we switch on
the perturbation by adding an asymmetric rotating component to the
magnetic field. We then let the system evolve during more than one
thousand orbital revolutions of the inner edge of the disk. We
performed four sets of simulations. In the first one, the
gravitational potential was Newtonian. In the second, we used a
pseudo-Newtonian potential in order to take the ISCO into account.
This is well-suited to describing the Schwarzschild space-time. In the
third one, we took the angular momentum of the star into account by
introducing a pseudo-Kerr geometry. And finally in the fourth and last
set, we performed simulations with a counter rotating accretion disk
evolving in the Newtonian potential described in the first set.

\subsection{Numerical test}

We have tested the pseudo-spectral method used in this paper on some
standard problems. For instance, we solved the Euler equations in
cylindrical coordinates for an initial jump in the density profile and
retrieved the classical results of the shock problem with the correct
shock speed.  However, in order to reduce the oscillations (Gibbs
phenomenon) due to discontinuity at the shock, we filtered out the
high frequency components in Fourier-Tchebyshev space.

In a second set of tests, we launched an~$m=1$ free wave in the
magnetized accretion disk, without any perturbation in the magnetic
field. The wave is slowly damped due to the absorbing boundary
conditions. The linear analysis elaborated in the previous section is
in agreement with the non-linear simulations presented below, at least
when the perturbations in the variables~$(\rho,r,\vec{v},\vec{B})$
remain small. For problems without shock or discontinuity in the flow,
this pseudo-spectral code gives very accurate solutions with only a few
discretization points.

\subsection{Newtonian potential}
\label{sec:ResNewt}

First, we study the behavior of the disk in the Newtonian potential.
In this case, the Keplerian rotation rate, the vertical and the radial
epicyclic frequencies for a single particle are all identical.
However, for a thin gaseous disk, there is a slight difference on the
order~$H/R$ and we have~:
\begin{equation}
\Omega_k \approx \kappa_r \approx \kappa_z
\end{equation}

The normalized rotation rate of the star around the $z$-axis
is~$\Omega_* = 0.0043311$. Assuming a~$1.4\,\mathrm{M_\odot}$ neutron
star, this corresponds to a spin of~$\nu_*=100$~Hz. The disk's inner
boundary is located at~$R_1=6$ and the outer boundary at~$R_2=60$. The
orbital angular motion at the inner edge of the disk
is~$\Omega_{in}=R_1^{-3/2}=0.0680$. We normalize the time by dividing
it by the spin period of the star~$T_* = \frac{2\,\pi}{\Omega_*} =
1450.7$.

\begin{figure}[h]
  \begin{center}
    \includegraphics[draft=false,scale=1]{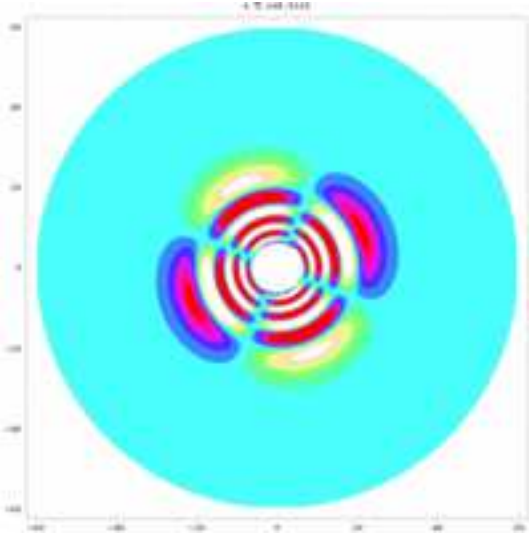}
    \caption{Final snapshot of the density perturbation~$\delta\rho/\rho_0$
      in the accretion disk evolving in a quadrupolar perturbed
      Newtonian potential.  The disk extends from~$R_1=6.0$
      to~$R_2=60.0$. The rotation rate of the star
      is~$\Omega_*=0.0043311$.  The time is normalized to the spin
      period~$T_*=1450.7$. The $m=2$ structure emerges in relation
      to the~$m=2$ quadrupolar magnetic perturbation.}
    \label{fig:DensDiscNewt} 
  \end{center}
\end{figure}

\begin{figure}[h]
  \begin{center}
    \includegraphics[draft=false,scale=1]{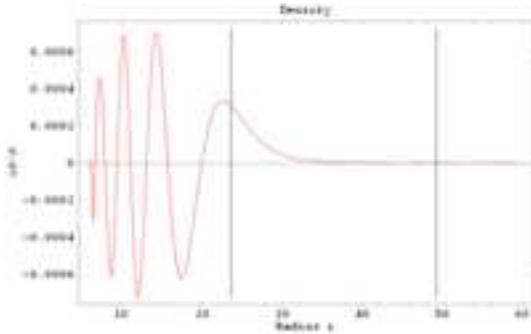}
    \caption{Cross section of the density perturbation~$\delta\rho/\rho_0$
      in the Newtonian disk at the final time of the simulation. The
      inner and outer Lindblad resonances are shown by vertical lines
      at~$r_L^{in/out}=23.7/49.3$. Nevertheless the disk feels only
      the inner resonance, the outer one being much too weak because
      of the very low perturbation amplitude in this region.}
    \label{fig:DensDiscNewt2} 
  \end{center}
\end{figure}

The final snapshot of the density perturbation in the disk calculated
as~$\Delta\rho/\rho_0=\rho/\rho_0-1$ is shown in
Fig.~\ref{fig:DensDiscNewt}. We clearly recognize the spatial
separation between the propagating and non-propagating regions
delimited by the inner Lindblad resonance located at~$R_L^{in}=23.7$.
However, the outer Lindblad resonance at~$R_L^{out}=49.3$ is not seen
in this run because the magnetic perturbation close to the outer
boundary is very weak compared to its amplitude close to the inner
edge. The corotation resonance, located in the neighborhood
of~$r=40.0$, is also not apparent. The reason is that its growth rate
is linear and very weak; an estimate is found by taking the typical
driven resonance for a harmonic oscillator as
\begin{equation}
  \label{eq:OscillateurForce}
  \frac{d^2 x}{dt^2} + \omega^2 \, x = f \, \cos (\omega\,t).
\end{equation}
which has a solution given by~:
\begin{equation}
  x(t) = \frac{f}{2\,\omega} \, t \, \sin(\omega\,t).
\end{equation}
The numerical value from the simulation is~$f/2\,\omega \approx
10^{-10}$. In order to get an appreciable amplitude at the corotation
point of~$10^{-3}$ of the average density, we would have to wait a
time~$\Delta t \ge 1500\,T_*$, which is much more than the total
duration of our run (about $70\,T_*$). This explains the missing
corotation resonance at this stage.  However, the most interesting
features are the very pronounced density fluctuations in the innermost
part of the accretion disk. They correspond to the non-wavelike
disturbance of the density related to the inhomogeneous solution to
the Schr\"odinger type equation~(\ref{eq:AppPsiSimp}).  They keep
their own identity even after 1000~revolutions of the innermost orbit
and rotate at the stellar spin rate.  After a few hundred orbital
revolutions, the disk settles down to a new quasi-stationary state in
which these disturbances persist.  This is confirmed by inspection of
Fig.~\ref{fig:DensDiscNewt2} in which we have plotted a cross section
of the density perturbation~$\delta\rho/\rho_0$ for a given azimuthal
angle, namely~$\varphi=0$.  As expected, the density perturbations
vanish at the disk edges. They behave as predicted by the linear
analysis in Fig.~\ref{fig:NonWave} when allowance is made for the
different boundary conditions.

\begin{figure}[h]
  \begin{center}
    \includegraphics[draft=false,scale=1]{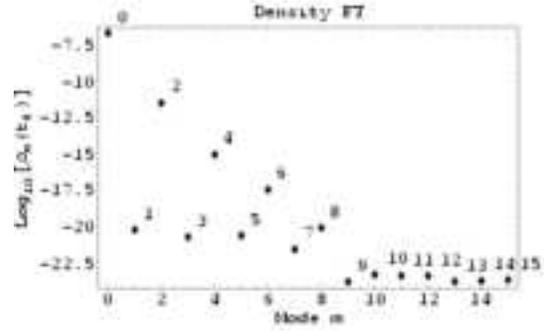}
    \caption{Amplitude of the Fourier components of the density perturbation. 
      The odd modes are numerically zero. Due to small nonlinearities,
      the even modes are apparent but with weak amplitude. The
      components~$m\ge9$ are set to zero because of the desaliasing
      process.}
    \label{fig:DensTFDiscNewt} 
  \end{center}
\end{figure}

The nonlinearities are weak throughout the simulation.  Indeed,
looking at the Fourier spectrum of the density in
Fig.~\ref{fig:DensTFDiscNewt}, the amplitude of each component,
plotted vs the mode~$m$ on a logarithmic scale, is small except for
that corresponding to the excitation mode.  The odd modes are not
present.  However, the weak nonlinearities create a cascade to high
even modes starting with~$m=2$. The largest asymmetric expansion
coefficient~$C_m$ is~$m=2$, while the next even coefficients roughly
follow a geometric series with a factor~$q=10^{-3}$, so we can write
for all~$m$ even, $C_m \approx q^{m/2-1} \, C_2$, until they reach
values less than~$10^{-20}$, which can be interpreted as zero from a
numerical point of view. As deviation from the stationary state is
weak, the amplitudes of these even modes decay compared to the
previous one, the highest being of course~$m=2$. Thus, even in the
full non-linear simulation, the regime remains quasi-linear.
Consequently, the parametric resonance phenomenon discussed in the
previous section is irrelevant at this stage of our work. The effects
of strong nonlinearity will be studied in a forthcoming paper. Note
that due to the desaliasing process, the modes~$m\ge9$ are all set to
zero.  Note also that the free wave solutions leave the computational
domain and do not return. Only the non-wavelike disturbance produces
significant changes in the density profile.

\subsection{Pseudo-Schwarzschild potential}
\label{sec:ResSchw}

In order to take into account the modification of the radial epicyclic
frequency due to the curved space-time around a Schwarzschild black
hole, we replaced the Newtonian potential by the logarithmically
modified potential proposed by Mukhopadhyay~(\cite{Mukhopadhyay}).
This potential is well-suited to approximating the angular and
epicyclic frequencies in accretion disks around a rotating black hole.
The radial gravitational field derived from this potential is given
by
\begin{equation}
  \label{eq:GravitePseudoGR}
  g_r = - \frac{G\,M_*}{r^2} \left[ 1 + R_{ms} \left( \frac{9}{20} \, 
      \frac{R_{ms}-1}{r} - \frac{3}{2r} \, \ln \frac{r}{(3r-R_{ms})^{2/9}} \right) \right],
\end{equation}
where~$R_{ms}$ is the last stable circular orbit~:
\begin{eqnarray}
  R_{ms} & = & 3+Z_2 \pm \sqrt{(3-Z_1)(3+Z_1+2Z_2)} \\
  Z_1 & = & 1 + (1-a^2)^{1/3} \, [ (1+a)^{1/3} + (1-a)^{1/3} ] \\
  Z_2 & = & \sqrt{3\,a^2 + Z_1^2},
\end{eqnarray}
where $a$ is related to the angular momentum of the star.

The important feature of this potential is that the radial epicyclic
frequency for a single particle vanishes at the innermost stable
circular orbit (ISCO), which implies that the parametric resonance
condition can be fulfilled in the very inner part of the accretion
disk for different values of the integer~$n$ but still with a very
small radial extension. However, due to the small perturbation we
remain in the linear regime, and the parametric resonance is only of
second order.  The disk boundaries, the star rotation rate, and the
normalization are the same as in the previous case.

\begin{figure}[h]
  \begin{center}
    \includegraphics[draft=false,scale=1]{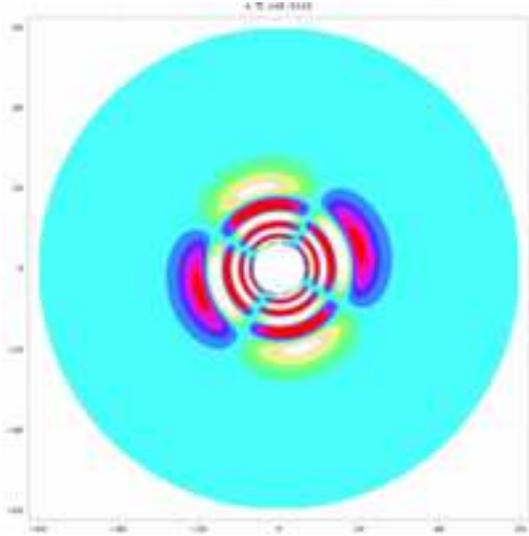}
    \caption{Same as in Fig.~\ref{fig:DensDiscNewt} but the
      Newtonian potential is replaced by a pseudo-Schwarzschild one.}
    \label{fig:DensDiscSchw} 
  \end{center}
\end{figure}

\begin{figure}[h]
  \begin{center}
    \includegraphics[draft=false,scale=1]{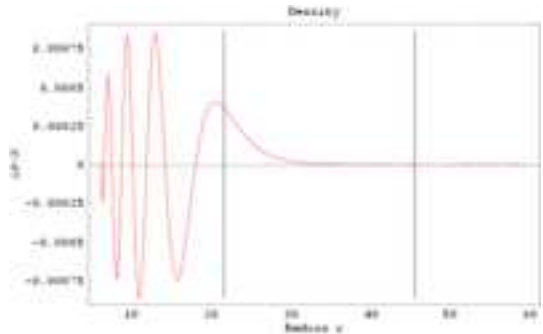}
  \end{center}
  \caption{Same as in Fig.~\ref{fig:DensDiscNewt2} but 
    for the pseudo-Schwarzschild potential. The Lindblad resonances
    are located at~$r_L^{in/out}=21.6/45.5$.}
   \label{fig:DensDiscSchw2} 
\end{figure}

\begin{figure}[h]
  \begin{center}
    \includegraphics[draft=false,scale=1]{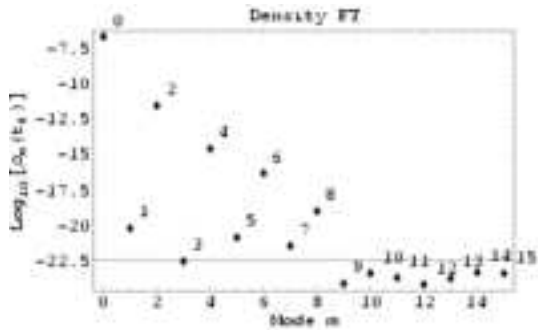}
    \caption{Same as Fig.~\ref{fig:DensTFDiscNewt} but 
      for the pseudo-Schwarzschild potential. The Fourier coefficients
      form a decaying geometric series.}
    \label{fig:DensTFDiscSchw} 
  \end{center}
\end{figure}

The final snapshot of the density perturbation in the disk is shown in
Fig.~\ref{fig:DensDiscSchw}. The corotation resonance is now located
at~$r=43.7$, which differs slightly from the previous simulation
because the orbital velocity is no longer Keplerian due to the
pseudo-Newtonian force Eq.~(\ref{eq:GravitePseudoGR}). The
non-wavelike disturbance exists between the inner edge of the disk and
the inner Lindblad resonance. By viewing a cross section, the Lindblad
radii appear clearly as seen in Fig.~\ref{fig:DensDiscSchw2}.
Inspection of the Fourier coefficients of the density
Fig.~\ref{fig:DensTFDiscSchw} proves that the linear regime is
preserved. The level of excitation of the even modes~$m\ne2$ is very
weak.

\begin{figure}[h]
  \begin{center}
    \includegraphics[draft=false,scale=1]{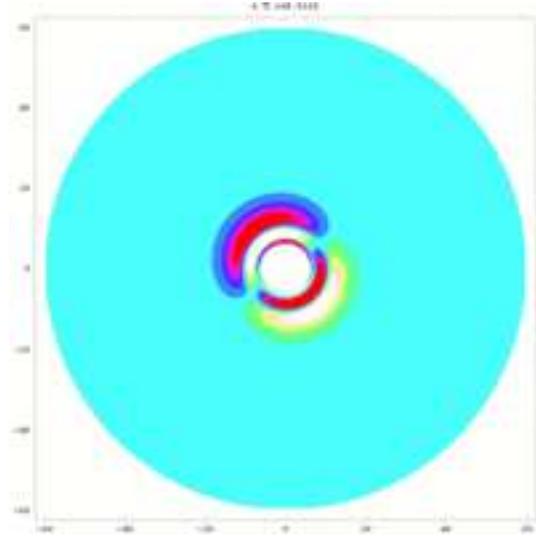}
    \caption{Same as in Fig.~\ref{fig:DensDiscSchw} but with an 
      azimuthal mode~$m=1$.}
    \label{fig:DensDiscSchwM1} 
  \end{center}
\end{figure}

\begin{figure}[h]
  \begin{center}
    \includegraphics[draft=false,scale=1]{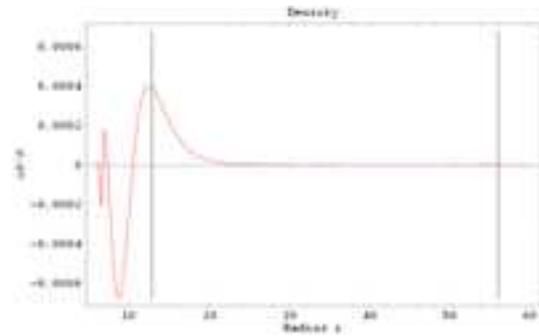}
  \end{center}
  \caption{Same as in Fig.~\ref{fig:DensDiscSchw2} but 
    with an azimuthal mode~$m=1$. The Lindblad resonances are located
    at~$r_L^{in/out}=12.9/56.0$.}
   \label{fig:DensDiscSchw2M1} 
\end{figure}

\begin{figure}[h]
  \begin{center}
    \includegraphics[draft=false,scale=1]{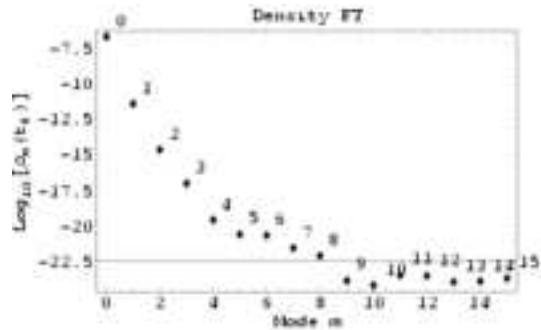}
    \caption{Same as Fig.~\ref{fig:DensTFDiscSchw} but 
      with an azimuthal mode~$m=1$. The Fourier coefficients form a
      decaying geometric series.}
    \label{fig:DensTFDiscSchwM1} 
  \end{center}
\end{figure}

The results of replacing the~$m=2$ mode by a more realistic~$m=1$
mode, in accordance with the dipolar structure of the magnetic field,
is shown in Fig.~\ref{fig:DensDiscSchwM1}.
The inner Lindblad resonance moves closer to the inner edge and is
located at~$R_L^{in}=12.9$. There is less space left for the
non-wavelike disturbance between the inner edge and the inner Lindblad
resonance as can be seen in Fig.~\ref{fig:DensDiscSchw2M1}.
Also, all the azimuthal modes are now excited starting with~$m=1$ and
with decreasing magnitude, Fig~\ref{fig:DensTFDiscSchwM1}.
From a qualitative point of view, we can draw the same conclusions as
in the case of the Newtonian disk. The absence or presence of the ISCO
does not affect the physics of the resonances, but only has an effect
on their amplitude and their location.

\subsection{Pseudo-Kerr potential}
\label{sec:ResKerr}

We also ran a simulation in which the rotation of the star is taken
into account. This shifts the location of the ISCO closer to the
surface of the neutron star.

We chose a star with an angular momentum of~$a_*=0.5$. Therefore, the
disk inner boundary corresponding to the marginally stable circular
orbit will be located at~$R_1=4.24$, while the outer boundary is chosen
to be at~$R_2=42.4$. The star magnetic perturbation in this run is ten
times larger than in the previous case, $\delta b = 0.01$. Therefore
the computation starts to leave the linear regime and reaches a weak
non-linear state.

\begin{figure}[h]
  \begin{center}
    \includegraphics[draft=false,scale=1]{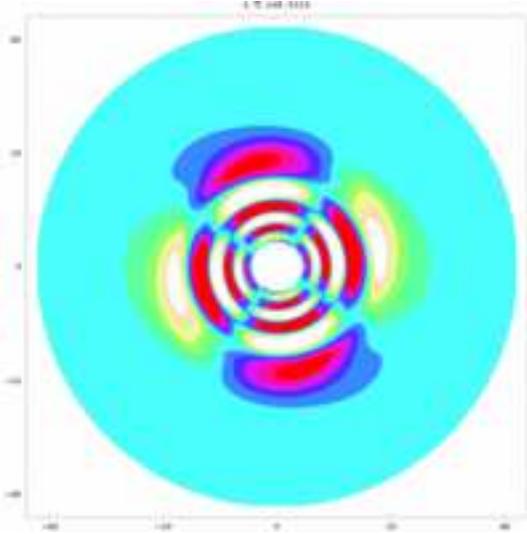}
    \caption{Final snapshot of the density perturbation in the accretion disk
      evolving in a perturbed pseudo-Kerr potential with~$a=0.5$. The
      disk extends from~$R_1=4.24$ to~$R_2=42.4$. The outer Lindblad
      resonance is not on the grid.}
    \label{fig:DensDiscKerr} 
  \end{center}
\end{figure}

\begin{figure}[h]
  \begin{center}
    \includegraphics[draft=false,scale=1]{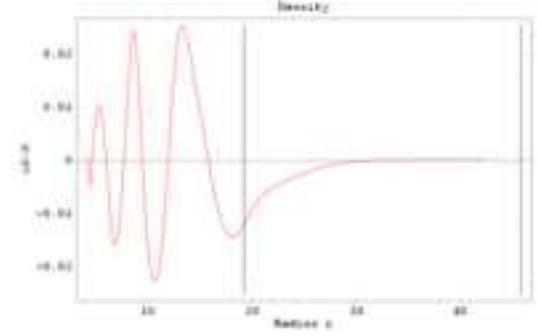}
    \caption{Same as in Fig.~\ref{fig:DensDiscNewt2} but 
      for the pseudo-Kerr potential. The inner Lindblad resonance
      appears clearly at~$r_L^{in}=19.2$ while the outer one
      at~$r_L^{out}=45.8$ lies outside the computational grid}
    \label{fig:DensDiscKerr2} 
  \end{center}
\end{figure}

\begin{figure}[h]
  \begin{center}
    \includegraphics[draft=false,scale=1]{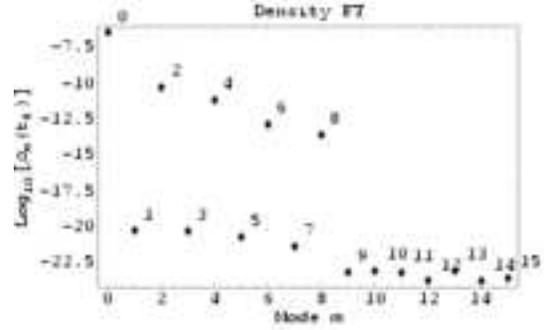}
    \caption{Same as Fig.~\ref{fig:DensTFDiscNewt} but
      for the pseudo-Kerr potential.}
    \label{fig:DensTFDiscKerr} 
  \end{center}
\end{figure}

The corotation resonance at~$r=23.8$ is not reached, as can be seen
from Fig.~\ref{fig:DensDiscKerr}. However, the wave and non-wavelike
disturbances are clearly identified between the inner edge and the
inner Lindblad resonance. They persist during the whole simulation at
the low level predicted by the linear analysis.  The precise location
of the Lindblad resonance is determined by the cross section view
shown in Fig.~\ref{fig:DensDiscKerr2}. Looking at the Fourier
components in Fig.~\ref{fig:DensTFDiscKerr}, all the even modes are
excited at a substantial level. The stronger stellar perturbation
starts to excite modes which are multiples of the injection scale
given by~$m=2$.

\subsection{Retrograde disk}

Finally we checked that the Lindblad resonances disappear for a
retrograde Newtonian disk. The numerical values are the same as in
Sec.~\ref{sec:ResNewt}, except that the sign of the stellar spin is
taken to be~$\Omega_*=-0.0043311$ and the magnetic
perturbation~$\delta b=0.01$.

\begin{figure}[h]
  \begin{center}
    \includegraphics[draft=false,scale=1]{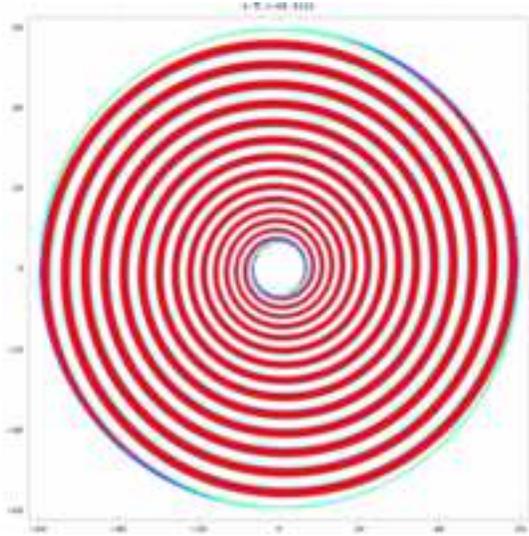}
    \caption{Final snapshot of the density perturbation in 
      the counterrotating accretion disk evolving in a perturbed
      Newtonian potential. Same values as in
      Fig.~\ref{fig:DensDiscNewt} apply except for the sign
      of~$\Omega_*$. A trailing spiral density wave of~$m=2$
      propagates in the entire disk.}
    \label{fig:DensDiscRetro} 
  \end{center}
\end{figure}

\begin{figure}[h]
  \begin{center}
    \includegraphics[draft=false,scale=1]{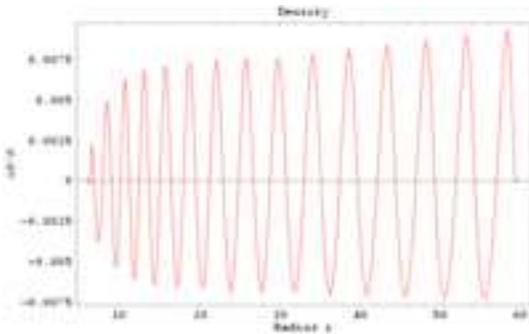}
    \caption{Same as Fig.~\ref{fig:DensDiscNewt2} but 
      for the retrograde disk. The Lindblad resonances are no longer
      present.}
    \label{fig:DensDiscRetro2} 
  \end{center}
\end{figure}

\begin{figure}[h]
  \begin{center}
    \includegraphics[draft=false,scale=1]{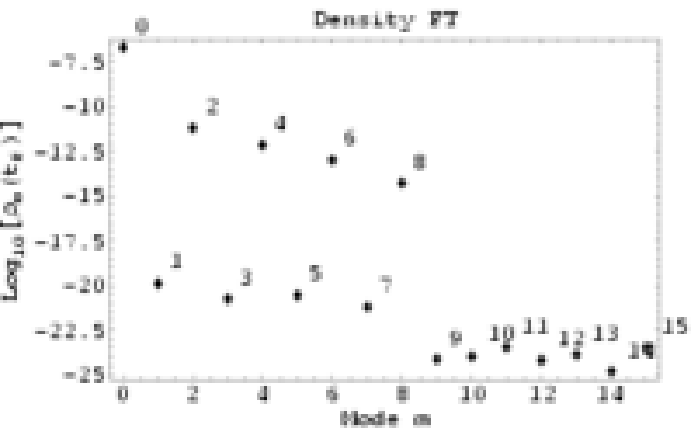}
     \caption{Same as Fig.~\ref{fig:DensTFDiscNewt} but 
      with a retrograde disk.}
    \label{fig:DensTFDiscRetro} 
  \end{center}
\end{figure}

Looking at Fig.~\ref{fig:DensDiscRetro}, the evolution of the disk
does not show any Lindblad resonance as expected. The mode~$m=2$ with
two spiral arms propagates in the whole computational domain.  The
disk reaches a quasi-stationary state in which the density
fluctuations are well established, Fig.~\ref{fig:DensDiscRetro2}.
They fit the shape found in the linear analysis,
Fig.~\ref{fig:NonWaveContre}. Again, the stronger stellar magnetic
perturbation affects the high order azimuthal modes, which are excited
at the same amplitude as the injection mode~$m=2$,
Fig.~\ref{fig:DensTFDiscRetro}.

\begin{figure}[h]
  \begin{center}
    \includegraphics[draft=false,scale=1]{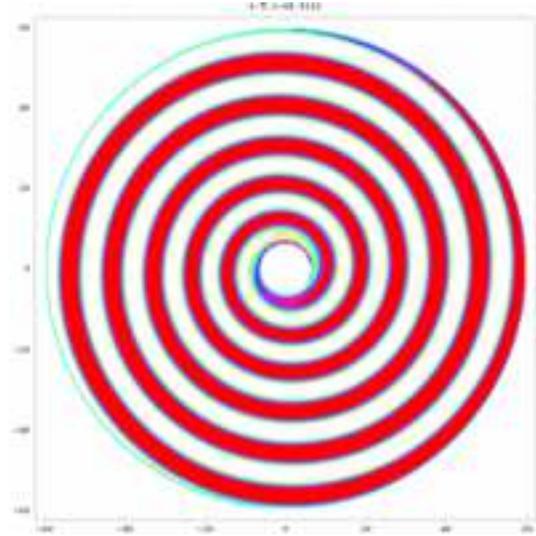}
    \caption{Final snapshot of the density perturbation in 
      the counterrotating accretion disk evolving in a perturbed
      Newtonian potential. Same values as in the
      Fig.~\ref{fig:DensDiscNewt} caption excepted for the sign
      of~$\Omega_*$. A trailing spiral density wave of~$m=1$ is
      propagating in the whole disk.}
    \label{fig:DensDiscRetroM1} 
  \end{center}
\end{figure}

\begin{figure}[h]
  \begin{center}
    \includegraphics[draft=false,scale=1]{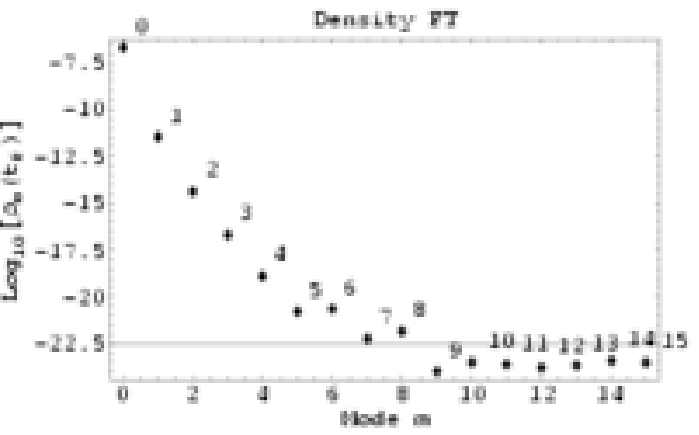}
     \caption{Same as Fig.~\ref{fig:DensTFDiscNewt} but 
       with a retrograde disk.}
    \label{fig:DensTFDiscRetroM1} 
  \end{center}
\end{figure}

Here also we can replace the~$m=2$ mode by a more realistic~$m=1$ one,
in accordance with the dipolar structure of the magnetic field and
with, for example, a smaller magnetic perturbation taken again to
be~$\delta b = 0.001$. The results for the density perturbation are
shown in Fig.~\ref{fig:DensDiscRetroM1}. A spiral wave is formed and
propagates in the whole disk with a rotation rate equal to the neutron
star spin.
\begin{figure}[h]
  \begin{center}
    \includegraphics[draft=false,scale=1]{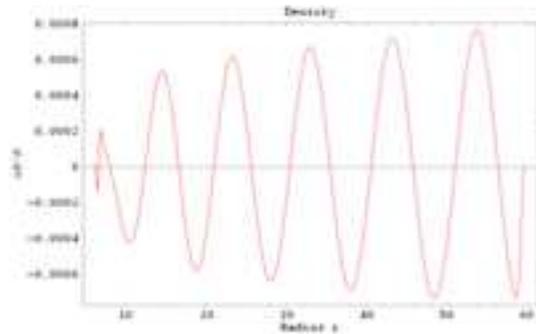}
    \caption{Same as Fig.~\ref{fig:DensDiscNewt2} but 
      with a retrograde disk. The Lindblad resonances have
      disappeared.}
    \label{fig:DensDiscRetro2M1} 
  \end{center}
\end{figure}
Now all the Fourier components are excited, the strongest one
corresponds to~$m=1$, and the lighter order ones have decreasing
amplitudes, as shown in Fig.~\ref{fig:DensTFDiscRetroM1}.

To summarize, in all of the results obtained by these runs, the exact
structure of the gravitational potential, be it Newtonian or
pseudo-Newtonian, does not affect the physics of the resonances.  The
main features of these resonances are their location very close to the
inner boundary of the accretion disk.  When the applied magnetic
asymmetry is small enough, the perturbations remain in the linear
regime.

\section{CONCLUSION}
\label{sec:Conclusion}

In this paper, we studied the effects of an asymmetric rotating
dipolar or multipolar magnetic field on a magnetized accretion disk.
This situation is well-suited to LMXBs containing a neutron star,
since they are believed to possess a magnetic field that more or less
channels the accreting matter to the polar caps of the neutron star,
depending on its strength. The same conclusions can be drawn in the
case of a hydrodynamical disk evolving in a quadrupolar gravitational
potential perturbation. Although the flows differ, the consequences of
a rotating asymmetric field are qualitatively the same.  The disk is
subject to the corotation resonance, the Lindblad driven resonances,
and if the perturbation is strong enough, the parametric instability.
We derived an explicit formulation for the regions where the blobs
should form, given by the resonance conditions
Eq.~(\ref{eq:ResCorot})-(\ref{eq:ResPara}). On the one hand, these
blobs can persist during thousands of orbital revolutions and can
account for the high quality factor~Q seen in some high
frequency-QPOs. On the other, the pattern speed of the free wave
solutions is closely related to the location of the inner boundary of
the accretion disk. The observed shift in the HFQPOs in response to
the accretion rate can be explained by these solutions.

This study should be extended to include viscosity of the gas flowing
in the disk, which should set a lower limit to the magnetic field
perturbation amplitude in order to have parametric resonance. In
addition, the loss of energy by radiation should also damp these
oscillations and select only a few of them, which will also permit
estimation of the light curves of such systems.  The relation between
the simulations presented in this paper and observations will be
investigated in a forthcoming paper, in which we show how to calculate
the power spectrum density (PSD) of the accretion disk.  The
prediction of the PSD will be crucial for validation of the model.

We would like to emphasize that this work is only a preliminary study
undertaken to prove that a rotating asymmetric magnetic field leads to
some interesting resonance mechanisms. The toy model used here cannot
be applied directly to observations. Indeed, many aspects, such as the
warping/precession of the disk expected in 3D simulations, as well as
viscosity and radiation, should be taken into account in a more
realistic model.

Nevertheless, we believe that this idea can also be applied to the
LMXBs containing a black hole. Indeed, it seems that the existence of
QPOs in black hole candidates is correlated with the detection of a
jet emanating from these systems, (McClintock \&
Remillard~\cite{MacClintock2003}). The Blandford-Znajek process
(Blandford \& Znajek~\cite{Blandford1977}) converting the rotational
energy of the black hole into the launching of the jet should also be
responsible for a connection between the accretion disk and the black
hole via the magnetic field. Therefore, the model presented in this
paper has the potential to explain the QPOs seen in the BHCs (Wang et
al.~\cite{Wang2003}).

\begin{acknowledgements}
  This research was carried out in an FOM projectruimte on
  `Magnetoseismology of accretion disks', a collaborative project
  between R. Keppens (FOM Institute Rijnhuizen, Nieuwegein) and N.
  Langer (Astronomical Institute Utrecht). This work is part of the
  research programme of the `Stichting voor Fundamenteel Onderzoek der
  Materie (FOM)', which is financially supported by the `Nederlandse
  Organisatie voor Wetenschappelijk Onderzoek (NWO)'.
  
  I am grateful to the referee Lev Titarchuk for his valuable comments
  and remarks as well as to Jean Heyvaerts and John Kirk for carefully
  reading the manuscript.
\end{acknowledgements}

%
%

\appendix


\section{Eigenvalue problem for the Lagrangian displacement}
\label{app:Eigenvalue}

\subsection{The Lagrangian displacement \vec{\xi}}

The linearization of the MHD~equations is done by introducing the
Lagrangian displacement vector~$\vec{\xi}$, which is related to the
Eulerian and Lagrangian perturbation of the fluid physical quantities,
such as the velocity, the magnetic field, and the density. These
perturbations are denoted respectively by~$\delta$ and~$\Delta$. By
definition, the Lagrangian perturbation of the velocity is
\begin{equation}
  \Delta\vec{v} = \vec{v} - \vec{v}_0 = \frac{d\vec{\xi}}{dt} 
  = \frac{\partial\vec{\xi}}{\partial t} + \vec{v}\cdot\grad \vec{\xi}
\end{equation}
Moreover, the Eulerian perturbation is related to the Lagrangian
one by
\begin{equation}
  \delta = \Delta - \vec{\xi}\cdot\nabla
\end{equation}
Therefore, the Eulerian perturbation of the velocity reads
\begin{equation}
  \label{appeq:PertbVit}
  \delta\vec{v} = \vec{v}_1 = \frac{\partial\vec{\xi}}{\partial t} + 
  \vec{v}\cdot\grad \vec{\xi} - \vec{\xi} \cdot \grad \vec{v} \\
\end{equation}
The boundary conditions at the disk-vacuum interface are
\begin{eqnarray}
  \label{eq:BC}
  < p + \frac{\vec{B}^2}{2\,\mu_0} > & = & 0 \\
  \vec{n} \cdot \vec{B} & = & 0 \\
  \vec{n} \wedge <\vec{B}> & = & 0
\end{eqnarray}
where~$<X>=X_{vac}-X_{disk}$ denotes the jump at the interface
and~$\vec{n}$ is a unit vector normal to the plasma interface.

Let's start by perturbing the mass conservation equation by using the
equilibrium condition~$\div(\rho_0\vec{v}_0)=0$
\begin{eqnarray}
  \frac{\partial\rho_1}{\partial t} + \vec{v}_0 \cdot \nabla\rho_1 + \rho_1 \, \div\vec{v}_0 
  = - \div(\rho_0 \, \vec{v}_1) =  & & \nonumber \\
  - \div\left( \rho_0 \, \frac{\partial\vec{\xi}}{\partial t} 
    + \rho_0 \, (\vec{v}\cdot\grad \vec{\xi} - \vec{\xi} \cdot \grad \vec{v}) \right)
\end{eqnarray}
From the tensorial calculus formalism, the right hand side can be
transformed after tedious algebra into
\begin{eqnarray}
  \frac{\partial\rho_1}{\partial t} + \vec{v}_0 \cdot \nabla\rho_1 + \rho_1 \, \div\vec{v}_0 
  = \frac{\partial}{\partial t} \left( -\div(\rho_0\,\vec{\xi}) \right) & & \nonumber \\
  + \vec{v}_0 \cdot \grad (-\div(\rho_0\,\vec{\xi})) - \div(\rho_0\,\vec{\xi}) \, \div\vec{v}_0
\end{eqnarray}
Because the Lagrangian displacement is arbitrary, we make the
following identification
\begin{equation}
  \rho_1 = -\div(\rho_0\,\vec{\xi})
\end{equation}
To continue by perturbing the induction equation, we get
\begin{eqnarray}
  \frac{\partial\vec{B}_1}{\partial t} + \vec{v}_0 \cdot \nabla \vec{B}_1 +
  \vec{B}_1 \, \div\vec{v}_0 - \vec{B}_1 \cdot \nabla \vec{v}_0 = & & \nonumber \\
  \vec{B}_0 \cdot \nabla \vec{v}_1 -
  \vec{B}_0 \, \div\vec{v}_1 - \vec{v}_1 \cdot \nabla \vec{B}_0 & &
\end{eqnarray}
We transform each term on the right hand side from the equilibrium
state $\rot(\vec{v}_0\wedge\vec{B}_0)=0$. The unperturbed magnetic
field is divergenceless neither is the perturbed one,
$\div\vec{B}_0=\div\vec{B}_1=0$. Thus, we obtain
\begin{eqnarray}
  \vec{B}_0 \cdot \nabla \vec{v}_1 = \frac{\partial}{\partial t} 
  (\vec{B}_0\cdot\nabla\vec{\xi})
  + \vec{v}_0\cdot\nabla(\vec{B}_0\cdot\nabla\vec{\xi}) + & & \nonumber \\
  \div\vec{v}_0 \, \vec{B}_0\cdot\nabla\vec{\xi} - 
  \div(\vec{B}_0\vec{\xi}\cdot\nabla\vec{v}_0 ) \\
  - \vec{B}_0 \, \div\vec{v}_1 = - \frac{\partial}{\partial t}(\vec{B}_0\div\vec{\xi}) 
  - \vec{v}_0\cdot\nabla(\vec{B}_0\div\vec{\xi}) +  & & \nonumber \\
  (\vec{v}_0\cdot\nabla\vec{B}_0)\div\vec{\xi}
  + \vec{B}_0 \,\vec{\xi}\nabla \div\vec{v}_0 \\
  - \vec{v}_1 \cdot \nabla \vec{B}_0 = - \frac{\partial}{\partial t} 
  (\vec{\xi}\cdot\nabla\vec{B}_0)
  - \vec{v}_0\cdot\nabla\vec{\xi}\cdot\nabla\vec{B}_0 +  & & \nonumber \\
  \vec{\xi}\cdot\nabla\left[ \vec{v}_0\cdot\nabla\vec{B}_0\right]
\end{eqnarray}
By introducing the vector
\begin{equation}
  \label{appeq:PertbMagnetic}
  \vec{Q}=\rot(\vec{\xi}\wedge\vec{B}_0) ,
\end{equation}
the sum of the three terms can be cast after tedious tensorial
calculus into the form
\begin{equation}
  \frac{\partial\vec{Q}}{\partial t} + \vec{v}_0\cdot\nabla\vec{Q} + 
  \vec{Q} \, \div\vec{v}_0 - \vec{Q} \cdot \nabla \vec{v}_0
\end{equation}
So we make the identification~$\vec{B}_1=\vec{Q}$. This is the
perturbation of the magnetic field induced by motion in the disk
and does not include the magnetic field perturbation emanating from
the star.

Finally for the pressure perturbation, we develop the adiabatic
condition to first order. We have
\begin{equation}
  \frac{\partial p_1}{\partial t} + \vec{v}_0 \cdot \nabla p_1 + 
  \gamma \, p_1 \, \div\vec{v}_0 = - \gamma \, p_0 \, \div\vec{v}_1 - 
  \vec{v}_1 \cdot \nabla p_0
\end{equation}
Introducing $\pi=-\gamma\,p_0\,\div\vec{\xi}-\vec{\xi}\cdot\grad p_0$,
the right hand side is
\begin{equation}
  \frac{\partial \pi}{\partial t} + \vec{v}_0 \cdot \nabla \pi + 
  \gamma \, \pi \, \div\vec{v}_0
\end{equation}
and thus the pressure perturbation of the gas is
\begin{equation}
  \label{appeq:PertbPression}
  p_1=\delta p = -\gamma\,p_0\,\div\vec{\xi}-\vec{\xi}\cdot\grad p_0
\end{equation}
Equations~(\ref{appeq:PertbVit}), (\ref{appeq:PertbMagnetic}),
(\ref{appeq:PertbPression}) are the perturbed quantities induced by
the motion in the fluid.  However, the magnetic field perturbation
possesses another variation that is not imposed by the fluid but is
introduced independently by the neutron star.  This known
perturbation~$\delta \vec{B}_*$ has to be added to the momentum
equation. In the realistic case of a dipolar magnetic field, we should
expect~$\rot\vec{B}_*=\vec{0}$, as well
as~$\rot\delta\vec{B}_*=\vec{0}$. Nevertheless, in our 2D numerical
simulations, we do not keep this property for the stellar magnetic
field because in this particular geometry, we impose straight field
lines aligned along the rotational $z$-axis. Therefore, in order to
treat the problem in a self-consistent manner we keep this
non-vanishing curl term in the linear analysis already in order to
facilitate the connection linear analysis-numerical simulations.  To
get the eigenvalues system satisfied by~$\vec{\xi}$ we put
expressions~(\ref{appeq:PertbMagnetic}, \ref{appeq:PertbPression})
into the momentum equation~(\ref{eq:DiscMHDv}).

\subsection{PDE for $\vec{\xi}$}

Replacing all the perturbed quantities by their expression involving
the Lagrangian displacement and keeping only the first order term,
$\vec{\xi}$, satisfies a second order linear PDE given by
\begin{eqnarray}
  \rho_0 \, \frac{\partial\vec{v}_1}{\partial t} + \rho_0 \, \vec{v}_0
  \cdot \nabla \vec{v}_1 + \rho_0 \, \vec{v}_1 \cdot \nabla \vec{v}_0 + 
  \rho_1 \, \vec{v}_0 \cdot \nabla \vec{v}_0 = & & \nonumber \\
  \rho_0 \delta\vec{g} + \rho_1 \, ( \vec{g}_0 + \delta\vec{g} ) - \nabla p_1 & + & \nonumber \\
  \frac{1}{\mu_0} \, \left( \rot \vec{B}_0 \wedge (\vec{B}_1 + \delta\vec{B}_*) + 
    \rot \vec{B}_1 \wedge (\vec{B}_0 + \right. & & \nonumber \\
  \left. \delta\vec{B}_*) + \rot \delta\vec{B}_* \wedge (\vec{B}_0 + 
    \vec{B}_1 + \delta\vec{B}_*) \right)
  & & 
\end{eqnarray}
On the left hand side, we have for each term
\begin{eqnarray}
  \rho_0 \, \frac{\partial\vec{v}_1}{\partial t} & = & 
  \rho_0 \, \frac{\partial^2\vec{\xi}}{\partial t^2} + \rho_0 \, \vec{v}_0 \cdot \nabla
  \frac{\partial\vec{\xi}}{\partial t} - \rho_0 \, \frac{\partial\vec{\xi}}{\partial t} \,
  \cdot \nabla \vec{v}_0 \\
  \rho_0 \, \vec{v}_0 \cdot \nabla \vec{v}_1 & = & \rho_0 \, \vec{v}_0 \cdot \nabla
  \frac{\partial\vec{\xi}}{\partial t} + \rho_0 \, \vec{v}_0 \cdot \nabla
  ( \vec{v}_0 \cdot \nabla \vec{\xi} - \vec{\xi} \cdot \nabla \vec{v}_0 ) \nonumber \\
 & & \\
  \rho_0 \, \vec{v}_1 \cdot \nabla \vec{v}_0 & = & \rho_0 \, \frac{\partial\vec{\xi}}{\partial t} \,
  \cdot \nabla \vec{v}_0
  + \rho_0 \, \left( \vec{v}_0 \cdot \nabla \vec{\xi} \cdot \nabla \right) \vec{v}_0
  \nonumber \\
  & & - \rho_0 \, \left( \vec{\xi} \cdot \nabla \vec{v}_0 \cdot \nabla \right) \vec{v}_0 \\
  \rho_1 \, \vec{v}_0 \cdot \nabla \vec{v}_0 & = & - \div(\rho_0\,\vec{\xi}) \, 
  \vec{v}_0 \cdot \nabla \vec{v}_0
\end{eqnarray}
Finally, summing these four terms gives
\begin{eqnarray}
  \rho_0 \, \frac{\partial\vec{v}_1}{\partial t} + \rho_0 \, \vec{v}_0
  \cdot \nabla \vec{v}_1 + \rho_0 \, \vec{v}_1 \cdot \nabla \vec{v}_0 + 
  \rho_1 \, \vec{v}_0 \cdot \nabla \vec{v}_0 = & & \nonumber \\
  \rho_0 \, \frac{\partial^2\vec{\xi}}{\partial t^2}
  + 2 \, \rho_0 \, \vec{v}_0 \cdot \nabla \frac{\partial\vec{\xi}}{\partial t}
  + \div\left[ \rho_0 \, \vec{v}_0 ( \vec{v}_0 \cdot \nabla ) \vec{\xi} -
  \rho_0 \, \vec{\xi} ( \vec{v}_0 \cdot \nabla ) \vec{v}_0 \right] & & \nonumber \\
& & 
\end{eqnarray}
Recalling that~$\vec{Q}=\vec{B}_1$ and~$p_1=\pi$, we rewrite
\begin{eqnarray}
  \rot \vec{B}_0 \wedge \vec{Q} + \rot \vec{Q} \wedge \vec{B}_0 =
  - \nabla(\vec{B}_0\cdot\vec{Q}) + \vec{B}_0\cdot \nabla \vec{Q}
  + \vec{Q} \cdot \nabla \vec{B}_0 & & \nonumber \\
  & & 
\end{eqnarray}

Introducing the opposite of the total pressure variation, gaseous and
magnetic, by~$\Pi = \gamma \, p_0 \, \div\vec{\xi} + \vec{\xi} \cdot
\grad p_0 - \frac{1}{\mu_0} \, \vec{B}_0 \cdot \vec{Q}$, the
Lagrangian displacement for the fluid motion in the accretion disk
satisfies the following second order linear partial differential
equation
\begin{eqnarray}
  \label{appeq:PDEXiGen}
  \rho_0 \, \frac{\partial^2\vec{\xi}}{\partial t^2}
  + 2 \, \rho_0 \, \vec{v}_0 \cdot \nabla \frac{\partial\vec{\xi}}{\partial t}
  +  & & \nonumber \\ 
  \div\left[ \rho_0 \, \vec{v}_0 ( \vec{v}_0 \cdot \nabla ) \vec{\xi} -
    \rho_0 \, \vec{\xi} ( \vec{v}_0 \cdot \nabla ) \vec{v}_0 \right] -  & & \nonumber \\ 
  \nabla\Pi - \frac{1}{\mu_0} \, \vec{B}_0 \cdot \nabla \vec{Q} 
  - \frac{1}{\mu_0} \, \vec{Q} \cdot \nabla \vec{B}_0 + & & \nonumber \\ 
  \div ( \rho_0 \, \vec{\xi} ) ( \vec{g}_0 + \delta\vec{g} ) - \rho_0 \, \delta\vec{g} =
  & & \nonumber \\ 
  \frac{1}{\mu_0} \, \left[ \rot ( \vec{B}_0 + \vec{Q} + \delta \vec{B}_* ) \wedge \delta \vec{B}_* 
    + \right. & & \nonumber \\ \left. \rot \delta \vec{B}_* \wedge ( \vec{B}_0 + \vec{Q} ) \right]
  & & 
\end{eqnarray}
In the expression above, we allow for a perturbation in the
gravitational field induced by the star and denoted
by~$\delta\vec{g}$.  However, in the following analysis, we remove
this freedom and only keep a perturbation in the magnetic field.

\subsection{Projection onto the cylindrical coordinate system}

In the stationary state, the disk orbits around the compact object
without mass in- or outflow from the edges and without precession.
From now on, we drop the subscript~$0$ for the stationary quantities,
which should not lead to any confusion. Thus the velocity in the
axisymmetric stationary state is simply
\begin{equation}
  \label{appeq:VitEq}
  \vec{v} = v_\varphi \, \vec{e}_\varphi = r \, \Omega \, \vec{e}_\varphi
\end{equation}
We find the following expressions
\begin{eqnarray}
  ( \vec{v} \cdot \nabla ) \vec{v} & = & - r \, \Omega^2 \, \vec{e}_r \\
  ( \vec{v} \cdot \nabla ) \vec{\xi} & = & \Omega \,
  \left[ \left( \frac{\partial\xi_r}{\partial\varphi} - \xi_\varphi \right) \, \vec{e}_r +
    \left( \frac{\partial\xi_\varphi}{\partial\varphi} + \xi_r \right) \, \vec{e}_\varphi +
    \frac{\partial\xi_z}{\partial\varphi} \, \vec{e}_z \right] \nonumber \\
  & & \\
  ( \vec{\xi} \cdot \nabla ) \vec{v} & = & - \Omega \, \xi_\varphi \, \vec{e}_r + 
  \vec{\xi} \cdot \grad (r\,\Omega) \, \vec{e}_\varphi
\end{eqnarray}
We also give the explicit formulas for each term in
Eq.~(\ref{appeq:PDEXiGen}).
\begin{eqnarray}
  \rho \, \frac{\partial^2\vec{\xi}}{\partial t^2} = 
  \rho \, \frac{\partial^2\xi_r}{\partial t^2} \, \vec{e}_r +
  \rho \, \frac{\partial^2\xi_\varphi}{\partial t^2} \, \vec{e}_\varphi +
  \rho \, \frac{\partial^2\xi_z}{\partial t^2} \, \vec{e}_z & & \\
  2 \, \rho \, \vec{v} \cdot \nabla \frac{\partial\vec{\xi}}{\partial t} =
  2 \, \rho \, \Omega \frac{\partial}{\partial t} \left[ 
    \left( \frac{\partial\xi_r}{\partial\varphi} - \xi_\varphi \right) \, \vec{e}_r + \right. & & \nonumber \\
    \left.
    \left( \frac{\partial\xi_\varphi}{\partial\varphi} + \xi_r \right) \, \vec{e}_\varphi + 
    \frac{\partial\xi_z}{\partial\varphi} \, \vec{e}_z
  \right] & & \\
  \grad\Pi = \frac{\partial\Pi}{\partial r} \, \vec{e}_r + 
  \frac{1}{r} \, \frac{\partial\Pi}{\partial\varphi} \, \vec{e}_\varphi + 
  \frac{\partial\Pi}{\partial z} \, \vec{e}_z & & \\
  - \div( \rho \, \vec{\xi} \, (\vec{v} \cdot \vec{\nabla}) \vec{v}) = & & \nonumber \\
  \left[ r \, \Omega^2 \, \div(\rho\,\vec{\xi}) + 
    \rho \, \vec{\xi} \, \grad( r \, \Omega^2 ) \right]
  \, \vec{e}_r + 
  \rho \, \xi_\varphi \, \Omega^2 \, \vec{e}_\varphi & & \\
  \div(\rho \, \vec{v} \, ( \vec{v} \cdot \vec{\nabla} ) \vec{\xi}) = 
  \rho \, \Omega^2 \, \frac{\partial^2 \vec{\xi}}{\partial\varphi^2} & & \\
  = \rho \, \Omega^2 \, \left[ 
    \left( \frac{\partial^2 \xi_r}{\partial\varphi^2} - \xi_r - 2 \, 
      \frac{\partial\xi_\varphi}{\partial\varphi} \right) \, \vec{e}_r + \right.
  & & \nonumber \\
  \left. \left( \frac{\partial^2 \xi_\varphi}{\partial\varphi^2} - \xi_\varphi + 2 \, 
      \frac{\partial\xi_r}{\partial\varphi} \right) \, \vec{e}_\varphi + 
    \frac{\partial^2 \xi_z}{\partial\varphi^2} \, \vec{e}_z \right] & & \nonumber \\
  \vec{B} \cdot \grad \vec{Q} = & & \nonumber \\
  \left[ B_r \, \frac{\partial Q_r}{\partial r} +
    \frac{B_\varphi}{r} \, \frac{\partial Q_r}{\partial \varphi} - 
    \frac{B_\varphi\,Q_\varphi}{r} +
    B_z \, \frac{\partial Q_r}{\partial z} \right] \, \vec{e}_r & & \nonumber \\
  + \left[ B_r \, \frac{\partial Q_\varphi}{\partial r} +
    \frac{B_\varphi}{r} \, \frac{\partial Q_\varphi}{\partial \varphi} + 
    \frac{B_\varphi\,Q_r}{r} +
    B_z \, \frac{\partial Q_\varphi}{\partial z} \right] \, \vec{e}_\varphi & & \nonumber \\
  + \left[ B_r \, \frac{\partial Q_z}{\partial r} + 
    \frac{B_\varphi}{r} \, \frac{\partial Q_z}{\partial \varphi} +
    B_z \, \frac{\partial Q_z}{\partial z} \right] \, \vec{e}_z & & \\
  \vec{Q} \cdot \grad \vec{B} = & & \nonumber \\
\left[ Q_r \, \frac{\partial B_r}{\partial r} +
    \frac{Q_\varphi}{r} \, \frac{\partial B_r}{\partial \varphi} - 
    \frac{Q_\varphi\,B_\varphi}{r} +
    Q_z \, \frac{\partial B_r}{\partial z} \right] \, \vec{e}_r & & \nonumber \\
  + \left[ Q_r \, \frac{\partial B_\varphi}{\partial r} +
    \frac{Q_\varphi}{r} \, \frac{\partial B_\varphi}{\partial \varphi} + 
    \frac{Q_\varphi\,B_r}{r} +
    Q_z \, \frac{\partial B_\varphi}{\partial z} \right] \, \vec{e}_\varphi \nonumber \\
  + \left[ Q_r \, \frac{\partial B_z}{\partial r} + 
    \frac{Q_\varphi}{r} \, \frac{\partial B_z}{\partial \varphi} +
    Q_z \, \frac{\partial B_z}{\partial z} \right] \, \vec{e}_z & & \\
  \rho \, \delta\vec{g} = \rho \, ( \delta g_r \, \vec{e}_r +  
  \delta g_\varphi \, \vec{e}_\varphi +  \delta g_z \, \vec{e}_z ) & &  \\
  - \div (\rho\,\vec{\xi} ) \, ( \vec{g} + \delta\vec{g} ) =  & & \nonumber \\
  - \div (\rho\,\vec{\xi} ) \, [ ( g_r + \delta g_r ) \, \vec{e}_r +
  \delta g_\varphi \, \vec{e}_\varphi + ( g_z + \delta g_z ) \, \vec{e}_z ] & & \\
  \rot( \vec{B} + \vec{Q} + \delta \vec{B}_* ) \wedge \delta \vec{B}_* = & & \nonumber \\
  \left[ \delta B_*^z \, \left\{ \frac{\partial}{\partial z} 
      \left( B_r + Q_r + \delta B_*^r \right) - \right. \right. & & \nonumber \\
\left.\left. \frac{\partial}{\partial r} 
      \left( B_z + Q_z + \delta B_*^z \right) \right\}  - \right. \nonumber & & \\
  \left. \frac{\delta B_*^\varphi}{r} \, \left\{ \frac{\partial}{\partial r} 
      \left( r \, (  B_\varphi + Q_\varphi + \delta B_*^\varphi ) 
      \right) - \right. \right. & & \nonumber \\
\left.\left. \frac{\partial}{\partial\varphi} ( B_r + Q_r + \delta B_*^r ) \right\}
  \right] \, \vec{e}_r & & \nonumber \\
  \left[ \frac{\delta B_*^r}{r} \, \left\{ \frac{\partial}{\partial r} 
      \left( r \, (  B_\varphi + Q_\varphi + \delta B_*^\varphi ) \right) -
      \right. \right. & & \nonumber \\
\left.\left. \frac{\partial}{\partial\varphi} ( B_r + Q_r + \delta B_*^r ) \right\} -\right.  & &\nonumber \\
  \left.
    \delta B_*^z \left\{ \frac{1}{r} \frac{\partial}{\partial\varphi} 
      (  B_z + Q_z + \delta B_*^z ) - \right. \right. & & \nonumber \\
\left.\left. \frac{\partial}{\partial z} 
      (  B_\varphi + Q_\varphi + \delta B_*^\varphi )\right\} \right] \, \vec{e}_\varphi
   & & \nonumber \\
   \left[ \delta B_*^\varphi \left\{ \frac{1}{r} \frac{\partial}{\partial\varphi} 
      (  B_z + Q_z + \delta B_*^z ) - \right. \right. & & \nonumber \\
\left.\left. \frac{\partial}{\partial z} 
      (  B_\varphi + Q_\varphi + \delta B_*^\varphi )\right\} -\right. & & \nonumber \\
  \left.
    \delta B_*^r \, \left\{ \frac{\partial}{\partial z} 
      \left( B_r + Q_r + \delta B_*^r \right) - \right. \right. & & \nonumber \\
\left.\left. \frac{\partial}{\partial r} 
      \left( B_z + Q_z + \delta B_*^z \right) \right\} \right] \, \vec{e}_z & & \\
  \rot \delta\vec{B}_* \wedge ( \vec{B} + \vec{Q} ) =  & & \nonumber \\
  \left[
    (B_z + Q_z) \, \left( \frac{\partial\delta B_*^r}{\partial z} -
      \frac{\partial\delta B_*^z}{\partial r} \right) - \right. & & \nonumber \\
    \left.
    \frac{B_\varphi + Q_\varphi}{r} \, \left( \frac{\partial}{\partial r}(r\,\delta B_*^\varphi) -
      \frac{\partial\delta B_*^r}{\partial\varphi} \right) \right] \, \vec{e}_r & & \nonumber \\
  \left[ \frac{B_r + Q_r}{r} \, \left( \frac{\partial}{\partial r}(r\,\delta B_*^\varphi) -
      \frac{\partial\delta B_*^r}{\partial\varphi} \right) - \right. & & \nonumber \\
    \left. (B_z + Q_z) \, 
    \left( \frac{1}{r} \, \frac{\partial\delta B_*^z}{\partial\varphi} -
      \frac{\partial\delta B_*^\varphi}{\partial z} \right) \right] \, \vec{e}_\varphi
  & &  \nonumber \\
   \left[ ( B_\varphi + Q_\varphi ) \left( \frac{1}{r} \, 
      \frac{\partial\delta B_*^z}{\partial\varphi} - 
      \frac{\partial\delta B_*^\varphi}{\partial z} \right) - \right. & & \nonumber \\
    \left. 
    (B_r + Q_r) \left( \frac{\partial\delta B_*^r}{\partial z} -
      \frac{\partial\delta B_*^z}{\partial r} \right) \right] \, \vec{e}_z 
  & & 
\end{eqnarray}
The perturbations in the disk magnetic field induced by the fluid are
in cylindrical coordinates
\begin{eqnarray}
  Q_r & = & \frac{1}{r} \, \frac{\partial}{\partial\varphi} 
  \left( \xi_r \, B_\varphi - \xi_\varphi \, B_r \right) -
  \frac{\partial}{\partial z} \left( \xi_z \, B_r - \xi_r \, B_z \right) \\
  Q_\varphi & = & \frac{\partial}{\partial z} \left( \xi_\varphi \, B_z - 
    \xi_z \, B_\varphi \right) -
  \frac{\partial}{\partial r} \left( \xi_r \, B_\varphi - \xi_\varphi \, B_r \right) \\
  Q_z & = & \frac{1}{r} \, \frac{\partial}{\partial r} 
  \left( r \, ( \xi_z \, B_r - \xi_r \, B_z ) \right) -
  \frac{1}{r} \, \frac{\partial}{\partial\varphi} 
  \left( \xi_\varphi \, B_z - \xi_z \, B_\varphi \right) \nonumber \\
& & 
\end{eqnarray}

Projecting on the 3~axis, and taking the stationary condition into
account
\begin{equation}
  \rho \, \left ( r \, \Omega^2 + g_r \right) = \frac{\partial}{\partial r} \, \left( 
    p + \frac{B_z^2}{2\,\mu_0} \right)
\end{equation}
The radial component of the Lagrangian displacement satisfies
\begin{eqnarray}
  \label{appeq:XiR}
  \rho \, \frac{D^2\xi_r}{Dt^2} - 
  2 \, \rho \, \Omega \, \frac{D\xi_\varphi}{Dt} 
  - \frac{\partial\Pi}{\partial r} + \frac{1}{\rho} \, \frac{dp}{dr} \,
  \div (\rho\,\vec{\xi}) + & & \nonumber \\
  \rho \, r \, \vec{\xi} \cdot \grad (\Omega^2) - 
  \frac{1}{\mu_0} \, \left[ B_r \, \frac{\partial Q_r}{\partial r} +
    \frac{B_\varphi}{r} \, \frac{\partial Q_r}{\partial \varphi} - \right. & & \nonumber \\
\left.
    2 \, \frac{B_\varphi\,Q_\varphi}{r} +
    B_z \, \frac{\partial Q_r}{\partial z} + Q_r \, \frac{\partial B_r}{\partial r} +
    \frac{Q_\varphi}{r} \, \frac{\partial B_r}{\partial \varphi} +
    Q_z \, \frac{\partial B_r}{\partial z} \right] & = & \nonumber \\
  ( \rho - \div (\rho\,\vec{\xi}) ) \, \delta g_r + \frac{1}{\mu_0} \,
  \left[ \delta B_*^z \, \left\{ \frac{\partial}{\partial z} 
      \left( B_r + Q_r + \delta B_*^r \right) - \right. \right. & & \nonumber \\
\left.\left. \frac{\partial}{\partial r} 
      \left( B_z + Q_z + \delta B_*^z \right) \right\} \right. & - & \nonumber \\
  \left. \frac{\delta B_*^\varphi}{r} \, \left\{ \frac{\partial}{\partial r} 
      \left( r \, (  B_\varphi + Q_\varphi + \delta B_*^\varphi ) 
      \right) -\right. \right. & & \nonumber \\
\left.\left. \frac{\partial}{\partial\varphi} 
      ( B_r + Q_r + \delta B_*^r ) \right\} \right. & + & \nonumber \\
  \left. (B_z + Q_z) \, \left( \frac{\partial\delta B_*^r}{\partial z} -
      \frac{\partial\delta B_*^z}{\partial r} \right) - \right. & & \nonumber \\
\left.
    \frac{B_\varphi + Q_\varphi}{r} \, \left( \frac{\partial}{\partial r}(r\,\delta B_*^\varphi) - \frac{\partial\delta B_*^r}{\partial\varphi} \right) \right]  & & \nonumber \\
& &
\end{eqnarray}
The azimuthal component is given by
\begin{eqnarray}
  \label{appeq:XiPhi}
  \rho \, \frac{D^2\xi_\varphi}{Dt^2} +
  2 \, \rho \, \Omega \, \frac{D\xi_r}{Dt} -
  \frac{1}{r} \, \frac{\partial\Pi}{\partial\varphi} - &  & \nonumber \\
  \frac{1}{\mu_0} \, \left[ B_r \, \frac{\partial Q_\varphi}{\partial r} +
    \frac{B_\varphi}{r} \, \frac{\partial Q_\varphi}{\partial \varphi} + 
    \frac{B_\varphi\,Q_r}{r} + \right. & & \nonumber \\
  \left. B_z \, \frac{\partial Q_\varphi}{\partial z} +\right. &  & \nonumber \\
  \left.  Q_r \, \frac{\partial B_\varphi}{\partial r} +
    \frac{Q_\varphi}{r} \, \frac{\partial B_\varphi}{\partial\varphi} + 
    \frac{Q_\varphi\,B_r}{r} +
    Q_z \, \frac{\partial B_\varphi}{\partial z} \right] = & & \nonumber \\
  ( \rho - \div (\rho\,\vec{\xi}) ) \, \delta g_\varphi +  & & \nonumber \\
  \frac{1}{\mu_0} \, 
  \left[ \frac{\delta B_*^r}{r} \, \left\{ \frac{\partial}{\partial r} 
      \left( r \, (  B_\varphi + Q_\varphi + \delta B_*^\varphi ) \right) -  \right. \right. &  & \nonumber \\
  \left. \left.
      \frac{\partial}{\partial\varphi} ( B_r + Q_r + \delta B_*^r ) \right\} - \right. &  & \nonumber \\
  \left.     \delta B_*^z \left\{ \frac{1}{r} \frac{\partial}{\partial\varphi} 
      (  B_z + Q_z + \delta B_*^z ) - \right. \right. & & \nonumber \\
    \left. \left. \frac{\partial}{\partial z} 
      (  B_\varphi + Q_\varphi + \delta B_*^\varphi )\right\} \right. & & \nonumber \\
  \left.\frac{B_r + Q_r}{r} \, \left( \frac{\partial}{\partial r}(r\,\delta B_*^\varphi) -
      \frac{\partial\delta B_*^r}{\partial\varphi} \right) - \right. & & \nonumber \\
    \left. (B_z + Q_z) \, 
    \left( \frac{1}{r} \, \frac{\partial\delta B_*^z}{\partial\varphi} -
      \frac{\partial\delta B_*^\varphi}{\partial z} \right) \right] & & \nonumber \\
\end{eqnarray}
and finally the vertical component satisfies
\begin{eqnarray}
  \label{appeq:XiZ}
  \rho \, \frac{D^2\xi_z}{Dt^2} - \frac{\partial\Pi}{\partial z} -
  \frac{1}{\mu_0} \, \left[ B_r \, \frac{\partial Q_z}{\partial r} + 
    \frac{B_\varphi}{r} \, \frac{\partial Q_z}{\partial \varphi} +
    B_z \, \frac{\partial Q_z}{\partial z} + \right. \nonumber \\
    \left. Q_r \, \frac{\partial B_z}{\partial r} + 
    \frac{Q_\varphi}{r} \, \frac{\partial B_z}{\partial \varphi} +
    Q_z \, \frac{\partial B_z}{\partial z} \right] = & & \nonumber \\
  ( \rho - \div (\rho\,\vec{\xi}) ) \, \delta g_z - \div (\rho\,\vec{\xi}) \, g_z & + & \nonumber \\
  \frac{1}{\mu_0} \, \left[ \delta B_*^\varphi \left\{ \frac{1}{r} \frac{\partial}{\partial\varphi} 
      (  B_z + Q_z + \delta B_*^z ) - \right. \right. & & \nonumber \\
  \left. \left.  \frac{\partial}{\partial z} 
      (  B_\varphi + Q_\varphi + \delta B_*^\varphi )\right\} - \right. \nonumber \\
    \delta B_*^r \, \left\{ \frac{\partial}{\partial z} 
      \left( B_r + Q_r + \delta B_*^r \right) -\right. & & \nonumber \\
  \left.  \frac{\partial}{\partial r} 
      \left( B_z + Q_z + \delta B_*^z \right) \right\} & + & \nonumber \\
  \left.  ( B_\varphi + Q_\varphi ) \left( \frac{1}{r} \, 
      \frac{\partial\delta B_*^z}{\partial\varphi} -
      \frac{\partial\delta B_*^\varphi}{\partial z} \right) - \right. & & \nonumber \\
  \left. 
    (B_r + Q_r) \left( \frac{\partial\delta B_*^r}{\partial z} -
      \frac{\partial\delta B_*^z}{\partial r} \right) \right]  & & \nonumber \\
& &
\end{eqnarray}

Eq.~(\ref{appeq:XiR})-(\ref{appeq:XiPhi})-(\ref{appeq:XiZ}) are the fundamental
equations governing the perturbed motion in the disk due to a small
gravitational and magnetic field.

\subsection{Orbital and vertical decoupled oscillations}

To elucidate the meaning of these equations, we focus on the disk
response when motions in vertical direction and orbital plane are
decoupled. Let's start with the Lagrangian displacement contained in
the orbital plane of the disk by setting $\frac{\partial}{\partial z}
= 0$ and~$\xi_z=0$. We also assume that perturbations are axisymmetric
in order to avoid meaningless complications so that
$\frac{\partial}{\partial\varphi}=0$, corresponding to~$m=0$ modes.
Then the magnetic perturbations read
\begin{eqnarray}
  Q_r & = & 0 \\
  Q_\varphi & = & 0 \\
  Q_z & = & - \frac{1}{r} \, \frac{\partial}{\partial r} 
  \left( r \, B_z \, \xi_r \right)
\end{eqnarray}
and the opposite of the total pressure perturbations
\begin{equation}
  \Pi = \gamma \, p \, \frac{1}{r} \, \frac{\partial}{\partial r} 
  \left( r \, \xi_r \right) + \xi_r \, \frac{\partial p}{\partial r} -
  \frac{B_z \, Q_z}{\mu_0}
\end{equation}
Keeping only the leading terms in Eq.~(\ref{appeq:XiPhi}) we find a
simple relation between radial and azimuthal displacement, namely
\begin{equation}
  \label{appeq:XiRPhi}
  \frac{D^2\xi_\varphi}{Dt^2} + 2 \, \Omega \, \frac{D\xi_r}{Dt} 
  \approx 0
\end{equation}
This can be integrated directly. Dropping the constant of integration
we find
\begin{equation}
  \label{appeq:DxiPhiR}
  \frac{D\xi_\varphi}{Dt} \approx - 2 \, \Omega \, \xi_r
\end{equation}
After straightforward algebra and making use of the equilibrium
condition~Eq.(\ref{eq:Equilibre3Dr}) and the
relation~(\ref{appeq:DxiPhiR}), the radial Lagrangian
displacement~(\ref{appeq:XiR}) satisfies the following second order
partial differential equation
\begin{eqnarray}
  \label{appeq:PDEXiRindpt}
  \rho \, \frac{D^2\xi_r}{Dt^2} - \frac{\partial}{\partial r} \left[ 
    \left( \rho \, ( c_s^2 + c_{az}^2 ) + \frac{B_z\,\delta B_*^z}{\mu_0} \right)
    \, \frac{1}{r} \, \frac{\partial}{\partial r} \left( r \, \xi_r \right) \right] + 
   & & \nonumber \\\rho \, \kappa_r^2 \, \xi_r
  = - \frac{\partial}{\partial r} \left( \frac{{\delta B_*^z}^2}{2\,\mu_0} +
    \frac{B_z\,\delta B_*^z}{\mu_0} \right) + \frac{\partial}{\partial r}
  \left( \frac{\delta B_*^z}{\mu_0} \, \frac{\partial B_z}{\partial r} \, \xi_r \right)
 & & \nonumber \\
 & & 
\end{eqnarray}
$\kappa_r^2 = 4 \, \Omega^2 + r \, \frac{\partial\Omega^2}{\partial
  r}$ is the radial epicyclic frequency of the oscillations.  The
sound speed and the Alfven velocity associated with the vertical
magnetic field are respectively
\begin{eqnarray}
  c_s^2 & = & \frac{\gamma\,p}{\rho} \\
  c_{az}^2 & = & \frac{B_z^2}{\mu_0\,\rho}
\end{eqnarray}
In the vertical direction, we neglect the variation in
the~$(r,\varphi)$ plane implying that~$\frac{\partial}{\partial r} =
\frac{\partial}{\partial\varphi}=0$. We set
also~$(\xi_r,\xi_\varphi)=0$ such that the magnetic perturbations
are
\begin{eqnarray}
  Q_r & = & - \frac{\partial}{\partial z} \, \left( B_r \, \xi_z \right) \\
  Q_\varphi & = & 0 \\
  Q_z & = & 0
\end{eqnarray}
and the opposite of the total pressure perturbations
\begin{equation}
  \Pi = \gamma \, p \, \frac{\partial\xi_z}{\partial z} 
  + \xi_z \, \frac{\partial p}{\partial z} -
  \frac{B_r \, Q_r}{\mu_0}
\end{equation}
After straightforward algebra and making use of the equilibrium
condition~Eq.(\ref{eq:Equilibre3Dz}), the vertical Lagrangian
displacement~(\ref{appeq:XiZ}) satisfies the following second order
partial differential equation
\begin{eqnarray}
  \label{appeq:PDEXiZindpt}
   \rho \, \frac{D^2\xi_z}{Dt^2} - \frac{\partial}{\partial z} \left[ 
    \left( \rho \, ( c_s^2 + c_{ar}^2 ) + \frac{B_r\,\delta B_*^r}{\mu_0} \right)
    \frac{\partial\xi_z}{\partial z} \right] + 
  \rho \, \kappa_z^2 \, \xi_z & = & \nonumber \\
  \frac{\partial}{\partial z} \left( \frac{{\delta B_*^r}^2+{\delta B_*^\varphi}^2}{2\,\mu_0} +
    \frac{B_r\,\delta B_*^r}{\mu_0} \right) + \frac{\partial}{\partial z}
  \left( \frac{\delta B_*^r }{\mu_0} \, \frac{\partial B_r}{\partial z} \, \xi_z \right)
  & & \nonumber \\
 & & 
\end{eqnarray}
The Alfven velocity associated with the radial magnetic field~$c_{ar}$
is
\begin{eqnarray}
c_{ar}^2 & = & \frac{B_r^2}{\mu_0\,\rho}
\end{eqnarray}

\subsection{Derivation of the inner boundary condition Eq.~(\ref{eq:Boundary})}

In the numerical runs presented in this paper, we use non reflecting
boundary conditions which leads to a vanishing density perturbation at
the inner edge of the disk, namely~$\delta\rho=0$. In order to compare
the linear analysis with simulations, we adopt the same conditions in
the both cases. Neglecting the displacement in the vertical direction,
density perturbations are related to the Lagrangian displacement by
\begin{equation}
  \delta\rho = - \div(\rho\,\vec{\xi}) =
  - \frac{1}{r} \, \frac{\partial}{\partial r}
  ( r \, \rho \, \xi_r) - i \, \frac{m}{r} \, \rho \, \xi_\varphi
\end{equation}
To the lowest approximation given by~Eq.~(\ref{appeq:DxiPhiR}), the
radial and azimuthal displacement expanded by~Eq.~(\ref{eq:DvlptX})
are related by
\begin{equation}
  \tilde{\xi}_\varphi \approx - 2 \, i \, \frac{\Omega}{\omega} \, \tilde{\xi}_r
\end{equation}
Injecting into the density perturbation, we have at the inner edge of
the disk~$r=R_1$
\begin{equation}
  \label{eq:DrhoR1}
  - \delta\rho = \frac{\partial}{\partial r}( r \, \rho \, \tilde{\xi}_r) +
  2 \, m \, \frac{\Omega}{\omega} \, \rho \, \tilde{\xi}_r = 0
\end{equation}
We recast the terms in such a way to obtain
\begin{equation}
  \frac{\partial\tilde{\xi}_r}{\partial r} + \left( \frac{\partial}{\partial r}
    \ln(r\,\rho) + 2 \, m \, \frac{\Omega}{r\,\omega} \, \rho \right) \, \tilde{\xi}_r = 0
\end{equation}
evaluated at~$r=R_1$. This is the inner boundary condition given by
Eq.~(\ref{eq:Boundary}).


\section{Approximate solution to the Schr\"odinger type equation}
\label{app:AppSchroeApprox}

In this appendix, we recall a method to find approximate solutions to
the following Schr\"odinger type equation with a source term~$f(x)$
\begin{equation}
  \label{eq:AppSchroeSrc}
  y''(x) + p(x) \, y(x) = f(x)
\end{equation}
The functions~$f$ and~$p$ are continuous in the interval~$[a,b]$,
which includes the origin~$x=0$, ($a<0,b>0$).  Moreover, $p$ has one
pole in this interval corresponding to the point~$x=0$ where it
changes sign, $p(0)=0$ and $p'(0)\ne0$.

\subsection{Homogeneous solution}

We start to look for solutions to the homogeneous part of
Eq.~(\ref{eq:AppSchroeSrc}) when~$f(x)=0$.  The reference equation is
Airy's equation given by
\begin{equation}
  \label{eq:AppAiry}
  w''(x) - x \, w(x) = 0
\end{equation}
We then look for approximate solutions to~Eq.(\ref{eq:AppSchroeSrc})
which can be cast into the following form, see for example
Smirnov~(\cite{Smirnov1989})
\begin{equation}
  \label{eq:AppSolApprox}
  y(x) = A(x) \, w(\omega_1(x))
\end{equation}
$w$ is an exact solution of Airy's equation. The functions $A$ and
$\omega_1$ are chosen such as to satisfy the homogeneous part
of~Eq.(\ref{eq:AppSchroeSrc}) within a prescribed accuracy. The first
and second derivatives of~$y(x)$ are
\begin{eqnarray}
  \label{eq:AppSolaPProxDeriv}
  y'(x) & = & A'(x) \, w(\omega_1(x)) + A(x) \, \omega_1'(x) \, w'(\omega_1(x)) \\
  y''(x) & = & A''(x) \, w(\omega_1(x)) + 2 \, A'(x) \, \omega_1'(x) \, w'(\omega_1(x))
  + \nonumber \\
   & + & A(x) \, \left( \omega_1''(x) \, w'(\omega_1(x)) (\omega_1')^2(x) \, w''(\omega_1(x)) \right)
\end{eqnarray}
However, due to~Eq.~(\ref{eq:AppAiry}), $w''(\omega_1(x)) =
\omega_1(x) \, w(\omega_1(x))$ and therefore the second order
derivative can be transform into
\begin{equation}
  \label{eq:AppSolApproxDer2}
  y'' = ( A'' + A \, (\omega_1')^2 \, \omega_1 ) \, w(\omega_1) + 
  ( 2 \, A' \, \omega_1' + A \, \omega_1'' ) \, w'(\omega_1(x)) 
\end{equation}
Inserting this expression into Eq.~(\ref{eq:AppSchroeSrc}), (remember
that we look for homogeneous solutions with~$f(x)=0$), we get
\begin{equation}
  \label{eq:Appsol1}
  ( A'' + A \, (\omega_1')^2 \, \omega_1 + p \, A ) \, w(\omega_1) + 
  ( 2 \, A' \, \omega_1' + A \, \omega_1'' ) \, w'(\omega_1(x)) = 0
\end{equation}
The Airy functions~$w$ and~$w'$ being linearly independent, $A$ and
$\omega_1$ satisfy the following system
\begin{eqnarray}
  \label{eq:Appsolsys1}
  A'' + A \, (\omega_1')^2 \, \omega_1 + p \, A & = & 0 \\
  \label{eq:Appsolsys2}
  2 \, A' \, \omega_1' + A \, \omega_1'' & = & 0
\end{eqnarray}
Assuming that the amplitude~$A$ is positive, Eq.~(\ref{eq:Appsolsys2})
is solved by
\begin{eqnarray}
  \label{eq:AppAx}
  A(x) = \frac{K}{\sqrt{|\omega_1'(x)|}}
\end{eqnarray}
where $K$ is constant. Neglecting the second derivative~$A''$ which is
one order of magnitude less than the other terms in
Eq.~(\ref{eq:Appsolsys1}), it reduces to
\begin{equation}
  \label{eq:Appdiff}
  (\omega_1')^2 \, \omega_1 + p = 0
\end{equation}
This relation implies that the function~$\omega_1$ has to satisfy the
property~$p(x)\,\omega_1(x)<0$.  We have to distinguish the following
two cases.

\subsubsection{$p(x)\ge0$}

It follows that~$\omega_1(x)\le0$.  The solution of
Eq.~(\ref{eq:Appdiff}) is
\begin{equation}
  \label{eq:AppOmega_1sol1}
  \omega_1(x) = - \left[ \pm \frac{3}{2} \, \int_0^x 
  \sqrt{p(s)} \, ds \right]^{2/3}
\end{equation}
The plus sign applies if~$p(x)\ge0$ for~$x\ge0$, otherwise the minus
sign applies.

\subsubsection{$p(x)\le0$}

Then~$\omega_1(x)\ge0$.  The solution of Eq.~(\ref{eq:Appdiff}) is
\begin{equation}
  \label{eq:AppOmega_1sol2}
  \omega_1(x) = \left[ \pm \frac{3}{2} \, \int_0^x 
  \sqrt{-p(s)} \, ds \right]^{2/3}
\end{equation}
The minus sign applies if~$p(x)\le0$ for~$x\le0$, otherwise the plus
sign applies.

Note that the function~$\omega_1$ is strictly monotonic. $\omega_1'$
never changes sign and never vanishes. Consequently, the amplitude~$A$
given by Eq.~(\ref{eq:AppAx}) remains finite in the entire
interval~$[a,b]$.

\subsubsection{Conclusion}

Finally, two linearly independent solutions of the homogeneous part of
Eq.~(\ref{eq:AppSchroeSrc}) can be expressed with help on the Airy
functions of the first and the second kind, respectively $Ai$ and
$Bi$
\begin{eqnarray}
  \label{eq:AppSolIndpt}
  y_1(x) & = & \frac{1}{\sqrt{|\omega_1'(x)|}} \, Ai(\omega_1(x)) \\
  y_2(x) & = & \frac{1}{\sqrt{|\omega_1'(x)|}} \, Bi(\omega_1(x)) 
\end{eqnarray}
The associated Wronskian is given by
\begin{equation}
  \label{eq:AppWronskien}
  W(y_1,y_2) = y_1 \, y_2' - y_1' \, y_2 = \frac{\mathrm{sign}(\omega_1')}{\pi}
\end{equation}
It never vanish, proving that the two solutions remain linearly
independent in the entire interval~$[a,b]$.

\subsection{Inhomogeneous solutions}

Knowing the general solutions to the homogeneous part, a peculiar
solution to the full Eq.~(\ref{eq:AppSchroeSrc}) with source term is
given by the formula
\begin{equation}
  \label{eq:AppSolPart}
  y_p(x) = \int^x \frac{y_1(x) \, y_2(t) - y_1(t) \, y_2(x)}{W(y_1,y_2)} \, f(t) \, dt
\end{equation}
Consequently, the most general approximate analytical solution to our
initial problem~Eq.~(\ref{eq:AppSchroeSrc}) to the lowest order is
\begin{eqnarray}
  \label{eq:AppSolFin}
  y(x) = C_1 \, y_1(x) + C_2 \, y_2(x) + & & \nonumber \\
  \mathrm{sign}(\omega_1') \, \pi
  \int^x (y_1(x) \, y_2(t) - y_1(t) \, y_2(x) ) \, f(t) \, dt & & 
\end{eqnarray}
As usual, the constants~$C_1,C_2$ are defined by properly fitting the
imposed boundary conditions at~$x=a$ and~$x=b$. Applied to the
linearized MHD accretion disk, we look for solutions which remains
bounded for large radii. This gives the first boundary condition. The
second boundary condition is derived by an appropriate density or
pressure perturbation at the inner edge. (See end of the appendix)


\section{Derivation of the resonance conditions}
\label{App:Resonances}

\subsection{Corotation resonance}

The corotation resonance is defined by the radius~$r_c$
where~$\omega_*(r_c)=0$. Actually, this equation possesses two
solutions corresponding to~$\omega(r_c)=\pm\frac{m\,c(r_c)}{r_c}$.
The width of this region is of the order of the disk height~$O(H)$.
For very thin disks, this discrepancy can be neglected and the two
solutions merge together in an unique corotation radius given
by~$\omega(r_c)=0$.  In other words, we assume in this case
that~$\omega=\omega_*$.  However, in our numerical application, the
separation between the two corotation radii is large enough to be
resolved. Nevertheless, in the derivation of the corotation resonance
below, we will assume that corotation is achieved
when~$\omega(r_c)=m\,(\Omega_*-\Omega(r_c))=0$. For the detailed study
of both corotation, it can be shown that the leading term in the
equation satisfied by the radial displacement~Eq.~(\ref{eq:XiR}) is
given by
\begin{eqnarray}
  \label{eq:CorotDominant}
  c_{maz}^2 \, \left[ 1 + \frac{m^2\,c_{maz}^2}{r^2\,\omega_*^2} \right] \,
  \frac{\partial^2\xi_r}{\partial r^2} + & & \nonumber \\
  \frac{c_{maz}^2}{r} \, \left[ 
    \frac{\partial \ln\,(r\,\rho\,c_{maz}^2)}{\partial\ln\,r} + \frac{m^2\,c_{maz}^2}{r} \, 
    \frac{\partial}{\partial r}(\omega_*^{-2}) \right] \,
  \frac{\partial\xi_r}{\partial r} + & & \nonumber \\
  \left[ \omega^2 - \kappa_r^2 + 4 \, \Omega^2\, 
    \left( 1 - \frac{\omega^2}{\omega_*^2} \right) + 
    2\,\Omega\,\omega \, \frac{m\,c_{maz}^2}{r} \, 
    \frac{\partial}{\partial r}(\omega_*^{-2}) \right] \, \xi_r & & \nonumber \\
  = \frac{m^2\,c_{maz}^2}{\rho\,r^2} \, \frac{B_z\,\delta B_*^z}{\mu_0} \,
  \frac{\partial}{\partial r}\left( \omega_*^{-2} \right) & &\nonumber \\
 & & 
\end{eqnarray}
Indeed, it is obtained from Eq.~(\ref{eq:XiR}) by replacing the
convective derivative~$D/Dt$ by $-i\,\omega$, $\xi_\varphi$ by
Eq.~(\ref{eq:XiPhiSol}) and neglecting second order terms.  Keeping
only the leading divergent terms in the coefficients of the ordinary
differential equation~(\ref{eq:CorotDominant}), we obtain
\begin{eqnarray}
  \label{eq:XiRCorot2}
  \frac{m^2\,c_{maz}^4}{r^2\,\omega_*^2} \, \xi_r''(r) + \frac{m^2\,c_{maz}^4}{r^2} \, 
  \frac{d}{dr}\left(\frac{1}{\omega_*^2}\right) \, \xi_r'(r) + & & \nonumber \\
  2\, \frac{m\,c_{maz}^2\,\Omega\,\omega}{r} \, 
  \frac{d}{dr}\left(\frac{1}{\omega_*^2}\right) \, \xi_r(r) 
  = \frac{m^2\,c_{maz}^2}{\rho\,r^2} \, \frac{B_z\,\delta B_*^z}{\mu_0} \,
  \frac{d}{dr}\left(\frac{1}{\omega_*^2}\right) & &\nonumber \\
   & & 
\end{eqnarray}
We introduce the new independent variable
\begin{equation}
  \label{eq:rc2x}
  x = \frac{r-r_c}{r_c}
\end{equation}
Expanding~$\omega_*$ to the first order around the corotation
radius~$r_c$ we have
\begin{eqnarray}
  \omega_*(r) = \omega_*(r_c) + (r-r_c)\,\omega_*'(r_c) + o(r-r_c)
  =  & & \nonumber \\
  x\,r_c\,\omega_*'(r_c) + o(x) \approx \alpha \, x
\end{eqnarray}
To this approximation, we have to solve
\begin{eqnarray}
  \label{eq:diff}
  \xi_r''(x) - \frac{2}{x} \, \xi_r'(x) - 
  4 \, \frac{\Omega\,\omega\,r_c^2}{m\,c_{maz}^2\,x} \, \xi_r(x) = & & \nonumber \\
  - 2\,\frac{r_c}{\rho\,c_{maz}^2\,x} \, \frac{B_z(r_c)\,\delta B_*^z(r_c)}{\mu_0} 
\end{eqnarray}
This is of the form
\begin{equation}
  \label{eq:yODE}
  y''(x) - \frac{2}{x} \, y'(x) - \frac{b}{x} \, y(x) = \frac{d}{x}
\end{equation}
with~$b = 4 \, \frac{\Omega\,\omega\,r_c^2}{m\,c_{maz}^2\,x}\ge0$ because
$\omega\,x = m(\Omega_*-\Omega)\,(r-r_c)/r_c \ge0$ and~$d = -2 \,
\frac{r_c}{\rho\,c_{maz}^2\,x} \, \frac{B_z(r_c)\,\delta
  B_*^z(r_c)}{\mu_0}$.  Making the change of
variable~$t=2\,\sqrt{b\,x}$ and introducing the new unknown~$v(t)$
by~$y(t)=t^3\,v(t)$, $v(t)$ satisfies the modified Bessel equation of
order~3
\begin{equation}
  \label{eq:BesselModif}
  v''(t) + \frac{1}{t} \, v'(t) - ( 1 + \frac{9}{t^2} ) \, v(t) = 0
\end{equation}
This is solved by
\begin{equation}
  \label{eq:BesselModifSol}
  v(t) = C_1 \, I_{3}(t) + C_2 \, K_{3}(t)
\end{equation}
Thus, the complete most general solution to Eq.~(\ref{eq:diff}) for
which a particular solution is easily found to be a constant equal
to~$\xi_r^p(r) = \frac{m}{2\,\Omega\,\omega\,\rho} \,
\frac{B_z(r_c)\,\delta B_*^z(r_c)}{\mu_0}$, is
\begin{eqnarray}
  \label{eq:n}
  \xi_r(x) = C_1 \, x^{3/2} \, I_3(2\,\sqrt{b\,x}) + 
  C_2 \, x^{3/2} \, K_3(2\,\sqrt{b\,x}) & & \nonumber \\
  + \frac{m}{2\,\Omega\,\omega\,\rho} \, \frac{B_z(r_c)\,\delta B_*^z(r_c)}{\mu_0}
\end{eqnarray}
Finally, near the corotation radius, the Lagrangian displacement which
remains bounded needs~$C_1=0$
\begin{equation}
  \label{eq:XiRForce}
  \xi_r(x) = C_2 \, x^{3/2} \, K_3(2\,\sqrt{b\,x}) 
  + \frac{m}{2\,\Omega\,\omega\,\rho} \, \frac{B_z(r_c)\,\delta B_*^z(r_c)}{\mu_0}
\end{equation}
The density disturbance induced in the disk by the rotating
gravitational perturbation is then to the lowest order
\begin{eqnarray}
  \label{eq:drhorho}
  \frac{\delta\rho}{\rho} = - \frac{\div(\rho\,\vec{\xi})}{\rho} = 
  - \frac{1}{\rho\,r}\,\frac{d}{dr}(r\,\rho\,\xi_r)
  - \frac{m}{r\,\omega_*^2} \, & & \nonumber \\
  \left( - \frac{m}{r\,\rho} \, 
    \frac{B_z(r_c)\,\delta B_*^z(r_c)}{\mu_0} +
    2\,\Omega\,\omega\,\xi_r \right) & & 
\end{eqnarray}
The displacement~Eq.~(\ref{eq:XiRForce}) is continuous and
differentiable everywhere. Thus, the first term on the right hand side
has a finite value. The second term on the RHS needs a special
treatment.  Indeed, when~$r$ approaches~$r_c$ the numerator$( -
\frac{m}{r\,\rho} \, \frac{B_z(r_c)\,\delta B_*^z(r_c)}{\mu_0} +
2\,\Omega\,\omega\,\xi_r )$ and the denominator~$\omega_*^2$ vanish as
well, leading to an undetermined expression of the form~$0/0$. To find
the behavior near~$r_c$ we introduce the function~$f(r) = -
\frac{m}{r\,\rho} \, \frac{B_z(r)\,\delta B_*^z(r)}{\mu_0}$. We note
that near the corotation, $(f(r) + 2\,\Omega\,\omega\,\xi_r)$ behaves
as $f(r)-f(r_c) = f'(r_c)\,(r-r_c)$ with~$f'(r_c)\ne0$.  Thus we
conclude that it approaches zero as~$x$. Therefore the density
perturbations in the disk diverge as
\begin{equation}
  \label{eq:drhorhoapprox}
  \frac{\delta\rho}{\rho} \approx - \, \frac{m\,f'(r_c)}
  {\omega_*'(r_c)^2 \, r_c^2 \, x}
\end{equation}
The divergent term in the density perturbation
Eq.~(\ref{eq:drhorhoapprox}) tends to infinity as~$\frac{1}{x}$. This
result is consistent with the conclusions drawn by Goldreich \&
Tremaine~(\cite{Goldreich1979}) for a hydrodynamic disk without
self-gravity. We just need to replace the sound speed~$c_s^2$ by the
fast magnetoacoustic wave speed~$c_{maz}^2$ and the gaseous pressure
by the total (gaseous+magnetic) pressure.

\subsection{Lindblad and parametric resonance}

By choosing an appropriate origin of time~$t_0$, the generalized
Mathieu equation~(\ref{eq:Typ}) can be recast in the more convenient
form as follows
\begin{equation}
  \label{eq:EqMathieu}
  \xi''(t) + \left[ \kappa^2 + h \, \cos ( \gamma \, t ) 
  \right] \, \xi(t) = f \, \cos( \gamma \, t )  
\end{equation}
where we have introduced the comoving excitation frequency~$\gamma =
\Omega - \Omega_*$.  It is a second order linear ordinary differential
inhomogeneous equation for the variable~$\xi$. The homogeneous part
corresponds exactly to Mathieu equation. The related solutions known
as Mathieu functions are well studied in mathematical physics
textbooks, see for example~Morse \& Feshbach~(\cite{Morse1953}).
These authors show that the solution becomes unbounded when the
excitation frequency~$\gamma$ lies in a close neighborhood to a
multiple of twice the harmonic frequency~$\kappa$, namely~$n\,|\gamma|
\approx 2\,\kappa$ where~$n$ is an integer. The size of the
neighborhood depends on the amplitude of the modulation~$h$. Because
the largest growth rate is reached when the relations holds exactly,
we only keep this fastest exponential growing solution. This gives the
parametric resonance condition as exposed in Eq.~(\ref{eq:ResPara}).

When the amplitude of the modulation is weak compared to the driven
force~$h\ll f$, the parametric growth rate can be neglected. Indeed,
to this approximation, Eq.(\ref{eq:EqMathieu}) reduce to the well
known driven harmonic oscillator
\begin{equation}
  \label{eq:OscForce}
  \xi''(t) + \kappa^2 \, \xi(t) = f \, \cos( \gamma \, t )  
\end{equation}
Resonance occurs when~$|\gamma|=\kappa$. In that case the most general
solution reads
\begin{equation}
  \label{eq:Sol}
  \xi(t) = A \, \cos(\kappa\,t) + B \, \sin(\kappa\,t) + 
  \frac{f\,t}{2\,\kappa} \, \sin(\kappa\,t)
\end{equation}
While in the parametric resonance the growth is exponential, here it
is only linear with time. Therefore, in the early stage of the full
solution of Eq.~(\ref{eq:EqMathieu}), this latter form is a good
approximation. On a very long timescale, it converges asymptotically the
Mathieu function with exponential growth rate.

Note that the driven resonance (called Lindblad resonance in the main
text to refer to the same physical process arising in galactic dynamic
and described by Lindblad~(\cite{Lindblad1974})) is a special case of
the parametric resonance for~$n=2$. As a consequence, we could enclose
the driven resonance condition into the parametric one. Nevertheless,
in order to keep track of the difference between linear and
exponential growth rate, this distinction will remain throughout the
paper.  Moreover, in the numerical application shown in the
simulations, the parametric resonance does not appear because of the
weak magnetic field disturbance. The timescale of the growing
parametric mode excess by several orders of magnitude the duration of
the simulations. Only the Lindblad resonance is relevant in this case.

\end{document}